\documentclass{sig-alternate}
\newcommand{\ignore}[1]{}
\usepackage[pass]{geometry}
\usepackage{fancyhdr}
\usepackage[normalem]{ulem}
\usepackage[hyphens]{url}
\usepackage{hyperref}
\usepackage{color}
\usepackage{soul}

\usepackage{algorithm}
\usepackage{algorithmicx}
\usepackage{algpseudocode}
\usepackage{subfig}
\usepackage{epstopdf}
\usepackage{multirow}
\usepackage{tikz}
\usepackage{amsmath,amssymb,amsfonts}
\usepackage{enumitem}
\usepackage{graphicx}

\fancypagestyle{firstpage}{
  \fancyhf{}

  \fancyhead[C]{\normalsize{To appear in 2020 IEEE International Symposium on High Performance Computer Architecture (HPCA).}}
  \pagenumbering{arabic}
}

\newlength\szg
\newcommand\hquan[1]{%
	\settoheight\szg{#1}%
	\tikz[baseline]{\pgfmathparse{
			ifthenelse(#1 < 10, 1, ifthenelse(#1 < 100, 0.75, 0.5))
		}
		\let\hfs\pgfmathresult
		\filldraw (0,\szg/2) circle (\szg/2+0.35ex);
		\node[white] at (0,\szg/2) {\makebox[0em][c]{\scalebox{\hfs}[1]{\textbf{#1}}}};
}}

\newcommand*\circled[1]{\tikz[baseline=(char.base)]{
		\node[shape=circle,draw,inner sep=.1pt] (char) {#1};}}

\definecolor{hpca}{rgb}{0 0 0}

\title{Enabling Highly Efficient Capsule Networks Processing Through A PIM-Based Architecture Design}

\numberofauthors{6}
\author{
	\alignauthor Xingyao Zhang\\
	\affaddr{University of Houston}\\
	\affaddr{Houston, USA}\\
	\email{xzhang55@uh.edu}
	\alignauthor Shuaiwen Leon Song\\
	\affaddr{University of Sydney}\\
	\affaddr{Sydney, Australia}\\
	\email{leonangel991@gmail.com}
	\alignauthor Chenhao Xie\\
	\affaddr{Pacific Northwest National Laboratory}\\
	\affaddr{Richland, USA}\\
	\email{fenahuhu@gmail.com}
	\and
	\alignauthor Jing Wang\\
	\affaddr{Capital Normal University}\\
	\affaddr{Beijing, China}\\
	\email{jwang@cnu.edu.cn}
	\alignauthor Weigong Zhang\\
	\affaddr{Capital Normal University}\\
	\affaddr{Beijing, China}\\
	\email{5591@cnu.edu.cn}
	\alignauthor Xin Fu\\
	\affaddr{University of Houston}\\
	\affaddr{Houston, USA}\\
	\email{xfu8@central.uh.edu}
}

\begin{document}
\maketitle

\thispagestyle{firstpage}
\pagestyle{plain}


\begin{abstract}

In recent years, the CNNs have achieved great successes in the image processing tasks, e.g., image recognition and object detection. Unfortunately, traditional CNN's classification is found to be easily misled by increasingly complex image features due to the usage of pooling operations, hence unable to preserve accurate position and pose information of the objects. To address this challenge, a novel neural network structure called Capsule Network has been proposed, which introduces equivariance through capsules to significantly enhance the learning ability for image segmentation and object detection. Due to its requirement of performing a high volume of matrix operations, CapsNets have been generally accelerated on modern GPU platforms that provide highly optimized software library for common deep learning tasks. However, based on our performance characterization on modern GPUs, CapsNets exhibit low efficiency due to the special program and execution features of their routing procedure, including massive unshareable intermediate variables and intensive synchronizations, which are very difficult to optimize at software level. To address these challenges, we propose a hybrid computing architecture design named \textit{PIM-CapsNet}. It preserves GPU's on-chip computing capability for accelerating CNN types of layers in CapsNet, while pipelining with an off-chip in-memory acceleration solution that effectively tackles routing procedure's inefficiency by leveraging the processing-in-memory capability of today's 3D stacked memory. Using routing procedure's inherent parallellization feature, our design enables hierarchical improvements on CapsNet inference efficiency through minimizing data movement and maximizing parallel processing in memory. Evaluation results demonstrate that our proposed design can achieve substantial improvement on both performance and energy savings for CapsNet inference, with almost zero accuracy loss. The results also suggest good performance scalability in optimizing the routing procedure with increasing network size.

\end{abstract}

\vspace{-.5em}
\section{Introduction}\label{sec:intro}

\begin{figure}
	\centering
	\includegraphics[width = 0.45\textwidth,height=1.2in]{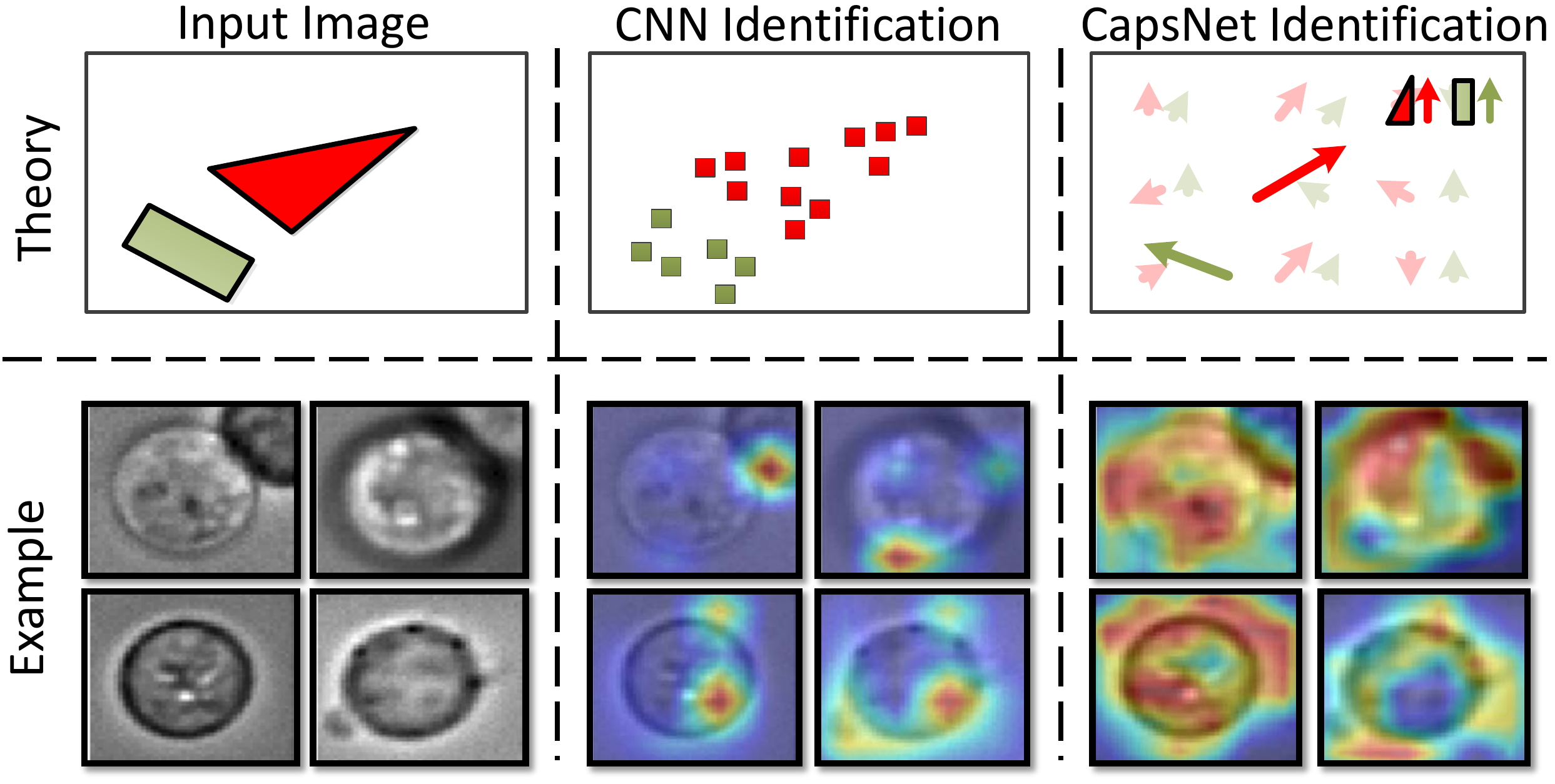}
	\vspace{-.5em}
	\caption{The comparison of neural networks in identifying lung cancer cells \cite{mobiny2018fast}, where CapsNet outperforms the traditional CNN on detection accuracy. The heat maps indicate the detected features.}
	\label{fig:exp}
	\vspace{-1.7em}
\end{figure}

Recently, machine learning has bloomed into rapid growth and been widely applied in many areas, including medical \cite{kononenko2001machine,erickson2017machine}, security \cite{buczak2016survey}, social media \cite{grimmer2015we}, engineering \cite{kober2013reinforcement} and etc.
Such explosive development is credited to the success of the neural network algorithms, especially the convolutional neural networks (CNNs), which are extremely suitable for the image processing tasks, e.g., image recognition and object detection \cite{ren2015faster,iandola2016squeezenet,szegedy2015going,simonyan2014very,he2016deep}.
\textcolor{black}{However, recent studies have found that CNNs could easily misdetect important image features when image scenarios are more complex. This could significantly affects the classification accuracy \cite{hinton2011transforming, hinton2014features, jaderberg2015spatial}. 
As shown in Fig.\ref{fig:exp}, when attempting to identify the lung cancer cells, traditional CNNs are bounded to the confined region, thus missing critical regions around the cell edges and leading to the wrong identification.}
\textcolor{black}{This is because the pooling operations used in common CNNs apply happenstance translational invariance \cite{hinton2011transforming,sabour2017dynamic} which limits the learning of rotation and proportional change, resulting in obtaining only partial features.} 
With the increasing adoption of emerging applications (e.g., \textcolor{black}{medical image processing} and autonomous driving) into humans' daily life that have strict requirements on object detection's accuracy, wrong identification could be fatal in some cases.
To address these challenges, a novel neural network called Capsule Network (CapsNet) has been proposed recently \cite{sabour2017dynamic}. 
Fig.\ref{fig:exp} demonstrates the evolution from classic CNN identification to CapsNet identification. CapsNet abandons the usage of pooling operations in CNNs and introduces the concept of \textit{capsule} which is any function that tries to predict the presence and the instantiation parameters of a particular object at a given location. The figure illustrates a group of capsules, each with a double-featured activation vector (i.e., probability of presence and pose, shown as the green and red arrows in Fig.\ref{fig:exp}). Because of this added equivariance, CapsNet can accurately detect the cancer cells via precise classification according to the cell edges and body texture. According to the recent studies, CapsNets are increasingly involved in the human-safety related tasks, and on average outperform CNNs by 19.6\% and 42.06\% on detection accuracy for medical image processing\cite{jimenez2018capsule,afshar2018capsule,mobiny2018fast,lundervold2018overview,zhang2019blood,wang2018sccapsnet} and autonomous driving\cite{kumar2018novel,kronenberger2018capsule,popperl2019capsule}.


Because CapsNets execution exhibits a high percentage of matrix operations, state-of-the-art GPUs have become primary platforms for accelerating CapsNets by leveraging their massive on-chip parallelism and deeply optimized software library \cite{lopez2018evolving,yusuf2018survey}. 
However, processing efficiency of CapsNets on GPUs often cannot achieve the desired level for fast real-time inference. To investigate the root causes of this inefficiency, we conduct a comprehensive performance characterization on CapsNets' execution behaviors on modern GPUs, and observe that the computation between two consecutive capsule layers, called \textit{routing procedure} (Sec.2.2), presents the major bottleneck. Through runtime profiling, we further identify that the inefficient execution of the routing procedure originates from (i) tremendous data access to off-chip memory due to the massive unshareable intermediate variables, and (ii) intensive synchronizations to avoid the potential write-after-read and write-after-write hazards on the limited on-chip storage. These challenges are induced by the unique features of the routing procedure execution, and cannot be addressed well via common NN optimization techniques \cite{stoutchinin2019optimally,li2016optimizing,song2017towards,chen2018multiple,zhang2018towards,chen2014dadiannao,du2015shidiannao,zhou2018cambricon,mahmoud2018diffy,kwon2018beyond,li2018network,deng2018permdnn} as well as software-level on-chip memory management (e.g., register manipulation or shared memory multiplexing).  





To tackle CapsNets' significant off chip-memory access and intensive synchronization (induced by numerous aggregation operations in the routing procedure), we propose a processing-in-memory based hybrid computing architecture named \textit{PIM-CapsNet}. At the highest level, PIM-CapsNet continues to utilize GPU's native on-chip units as the host for fast processing CNN-type of layers such as Convolution and Fully-connected included in CapsNet. Meanwhile, by leveraging batch execution, PIM-CapsNet pipelines the host GPU execution with an off-chip in-memory solution that can effectively accelerate CapsNet's routing procedure. 

To enable the in-memory acceleration capability for the routing procedure (RP), we select one of the emerging 3D stacked technologies, i.e., hybrid memory cube (HMC)\cite{jeddeloh2012hybrid}, for RP's design and integration. Because of HMC's high internal and external memory bandwidth, and its unique logic layer for easy integration of computation logic, it has become a promising PIM platform \cite{yitbarek2016exploring,ahn2015pim,ahn2016scalable,liu2018processing,kim2016neurocube}. By leveraging HMC's unique architectural features, our PIM-CapsNet design mitigates the data-intensive RP into HMC to tackle its major challenges above. There are two key objectives for this in-memory design for RP: under the hardware design constraints, creating a hierarchical optimization strategy to enable (i) inter-vault workload balance and communication minimization (Sec.5.1 and 5.3) and (ii) intra-vault maximum parallel processing (Sec.5.2 and 5.3). Additionally, the in-memory optimizations should be generally applicable to different RP algorithms.  

For objective (i), we further investigate RP's algorithm and identify an interesting feature: highly parallelizable in multi-dimensions. This makes RP's workloads have great potential to be concurrently executed across vaults without incurring significant communication overheads. We then create a modeling strategy by considering both per-vault workloads and inter-vault communication to guide dimension selection for parallelization, in order to achieve the optimal performance and power improvement. This also significantly reduces the original synchronization overheads through the conversion to aggregation within a vault. For objective (ii), we integrate multiple processing elements (PEs) into a vault's logic layer and explore the customized PE design to concurrently perform RP's specific operations across memory banks. Meanwhile, we propose a new address mapping scheme that effectively transfers many inter-vault level data requests to the intra-vault level and address the bank conflict issues of concurrent data requests. Furthermore, to reduce logic design complexity and guarantee performance, we use simple low-cost logic to approximate complex special functions with negligible accuracy loss. To summarize, this study makes the following contributions:
\begin{itemize}
	\vspace{-.5em}
	\item We conduct a comprehensive characterization study on CapsNet inference on modern GPUs and identify its root causes for execution inefficiency. 
	\vspace{-.7em}
	\item Based on the interesting insights from the characterization and further algorithm analysis, we propose a processing-in-memory based hybrid computing architecture named \textit{PIM-CapsNet}, which leverages both GPU on-chip computing capability and off-chip in-memory acceleration features of 3D stacked memory to improve the overall CapsNet inference performance.
	
	\vspace{-.7em}
	\item To drastically reduce the identified performance bottlenecks, we propose several memory-level optimizations to enable minimal in-memory communication, maximum parallelization, and low design complexity and overhead.  
	
	\vspace{-.7em}
	\item The experimental results demonstrate that for the overall CapsNet inference, our proposed \textit{PIM- CapsNet} design outperforms the baseline GPU by 2.44x (upto 2.76x) on performance and 64.91\% (upto 85.16\%) on energy saving. It also achieves good performance scalability with increasing network size.

	
\end{itemize}

\vspace{-1.2em}
\section{Background: Capsule Network}\label{sec:bg}

\begin{figure}
	\centering
	\includegraphics[width = 0.48\textwidth,height=.7in]{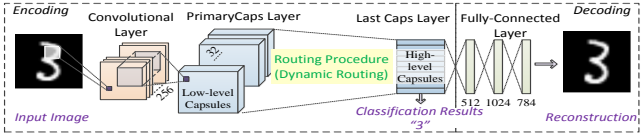}
	\vspace{-1.3em}
	\caption{The computation diagram of CapsNet for MNIST \cite{sabour2017dynamic}.}
	\label{fig:capsnet}
	\vspace{-1.5em}
\end{figure}

As shown in Fig.\ref{fig:capsnet}, CapsNet inherits the convolutional (Conv) and fully connected (FC) layers from the standard CNNs, but introduces new layers (i.e., Caps layers) to realize the concept of ``capsule" for better information representation. A capsule is a group of neurons (Fig.\ref{fig:exp}) whose activity vector represents instantiation parameters of a specific type entity (e.g., the location and pose of an object). It introduces equivariance which makes standard CNNs understand rotation and proportional change (e.g., in Fig.\ref{fig:exp}). CapsNet significantly lifts the limitation of  happenstance translational invariance of pooling operations applied in the traditional CNNs, thus being considered to be superior in image segmentation and object detection \cite{hinton2011transforming,sabour2017dynamic}. 

\vspace{-.6em}
\subsection{CapsNet Structure}\label{sec:caps_struct}

Fig.\ref{fig:capsnet} takes CapsNet-MNIST \cite{sabour2017dynamic} as an example to illustrate a basic CapsNet structure. It contains two computation stages: encoding and decoding. The encoding stage is composed of Conv layers and Caps layers. The convolutional operations are first performed on the data mapping from neurons of the Conv layer to the capsules of the first Caps layer, which is defined as PrimeCaps layer. It is often followed by at least one other Caps layer. The data mapping between the capsules of the adjacent Caps layers is performed via the \textit{routing procedure (RP)}. Finally, the last Caps layer produces the classification information towards categorization, with each capsule representing one category. Following the encoding stage, the decoding function includes multiple FC layers which attach to the last Caps layer for image reconstruction (i.e., improving model accuracy in training or plotting abstract features in inference) \cite{sabour2017dynamic}.

\begin{figure}
	\centering
	\includegraphics[width = 0.48\textwidth,height=1.4in]{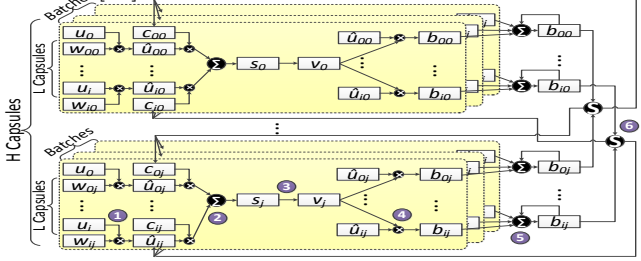}
	\vspace{-1.5em}
	\caption{Dynamic Routing Procedure (RP) in CapsNet: describing computation flow and multiple ways for parallel processing. }
	\vspace{-.5em}
	\label{fig:DR}
\end{figure}

\begin{algorithm}[!t]
	\scriptsize
	\caption{Dynamic Routing Procedure}	
	\label{ag:cap}
	\floatname{algorithm}{Procedure}
	\renewcommand{\algorithmicrequire}{\textbf{Input:}}
	\renewcommand{\algorithmicensure}{\textbf{Output:}}
	\begin{algorithmic}[1]
		\Require L capsules $u$, weight matrix $W$
		\Ensure H capsules $v$		
		
		\State for all L capsule $i$ \& all H capsule $j$ from all input set $k$:
		\Statex \qquad $\hat{u}^k_{j|i} \gets u^k_i \times W_{ij}$ \Comment{Eq.\ref{eq:uhat}}
		
		\State for all L capsule $i$ \& all H capsule $j$: 
		\Statex \qquad $b_{ij}\gets 0$ \Comment{Initialize Routing Coeffcients}
		\For{Routing iterations}
		\State for all L capsule $i$: 
		\Statex \qquad \qquad $c_{ij}\gets softmax(b_{ij})$ \Comment{Eq.\ref{eq:cij}}
		\State for all H capsule $j$ from all input set $k$: 
		\Statex \qquad \qquad $s^k_j \gets \sum_{i} \hat{u}^k_{j|i} \times c_{ij}$ \Comment{Eq.\ref{eq:sj}}
		\State for all H capsule $j$ from all input set $k$: 
		\Statex \qquad \qquad $v^k_j \gets squash(s^k_j)$ \Comment{Eq.\ref{eq:vj}}
		\State for all L capsule $i$ \& H capsule $j$: 
		\Statex \qquad \qquad $b_{ij}=\sum_{k}v^k_j\hat{u}^k_{j|i} + b_{ij}$ \Comment{Eq.\ref{eq:bij}}
		\EndFor
		\State \textbf{Return} $v$
	\end{algorithmic}
	\vspace{-.3em}
\end{algorithm}



\vspace{-.5em}
\subsection{Essential Mechanism For Avoiding Feature Loss: Routing Procedure (RP)}
\label{sec:2.2}

The \textit{routing procedure} (RP)
is introduced to route the information from low-level capsules (or \textit{L} capsules) to high-level capsules (or \textit{H} capsules) without feature loss. There have been several routing algorithms used in routing procedure such as Dynamic Routing \cite{sabour2017dynamic} and  Expectation-Maximization routing\cite{hinton2018matrix}. 
In this work, we use the popular Dynamic Routing \cite{sabour2017dynamic} as an example to explain the RP execution.

Fig.\ref{fig:DR} and Algorithm.\ref{ag:cap} demonstrate the computation flow and possible dimensions for parallelization. 
Given the $k$th batched input set, in order to generate $j$th H capsule, $i$th L capsule in this input set ($u^k_i$) first multiplies with the corresponding weight ($W_{ij}$) to generate the prediction vector ($\hat{u}_{j|i}$) (Fig.\ref{fig:DR} \hquan{1}):
\vspace{-.5em}
\begin{equation}
\resizebox{.23\hsize}{!}{$\hat{u}^k_{j|i}=u^k_i\times W_{ij}$}
\label{eq:uhat}
\vspace{-.5em}
\end{equation}
Then, these prediction vectors will multiply with their corresponding routing coefficients ($c_{ij}$) with results aggregated across all L capsules (Fig.\ref{fig:DR} \hquan{2}):
\vspace{-.7em}
\begin{equation}
\resizebox{.23\hsize}{!}{$s^k_j=\sum_{i}\hat{u}^k_{j|i}\times c_{ij}$}
\label{eq:sj}
\vspace{-.7em}
\end{equation}
The non-linear ``squashing" function is then implemented on the aggregated results of $s^k_j$ to produce the $j$th H capsule $v_j$ (Fig.\ref{fig:DR} \hquan{3}):
\vspace{-1em}
\begin{equation}
\resizebox{.24\hsize}{!}{$v^k_j=\frac{||s^k_j||^2}{1+||s^k_j||^2}\frac{s^k_j}{||s^k_j||}$}
\label{eq:vj}
\vspace{-.5em}
\end{equation}
Note that the $v^k_j$ can not be considered as the final value of $j$th H capsule unless the features of L capsules have been correctly inherited.
The information difference between an L and H capsule can be quantified by the agreement measurement via the scalar production of the prediction vector $\hat{u}^k_{j|i}$ and the H capsule $v^k_j$ (Fig.\ref{fig:DR} \hquan{4}), where the ``0'' output means the information is precisely inherited.
In the case of a large divergence, the agreements will be accumulated into an intermediate variable $b_{ij}$ (Fig.\ref{fig:DR} \hquan{5}), which will be used to update the routing coefficients via the ``softmax'' function (Fig.\ref{fig:DR} \hquan{6}).
\vspace{-.5em}
\begin{equation}
\resizebox{.26\hsize}{!}{$b_{ij}=\sum_{k}v^k_j\hat{u}^k_{j|i} + b_{ij}$}
\label{eq:bij}
\vspace{-.5em}
\end{equation}
\vspace{-1em}
\begin{equation}
\resizebox{.25\hsize}{!}{$c_{ij}=\frac{exp(b_{ij})}{\sum_{k}exp(b_{ik})}$}
\label{eq:cij}
\vspace{-.3em}
\end{equation}
The updated routing coefficients will then be integrated in Eq.(\ref{eq:sj}) to start another iteration in this routing procedure (Fig.\ref{fig:DR} \hquan{2}). 

The number of iterations is determined by the convergence of routing coefficients and set by programmers. Several recent studies indicate that the number of iterations increases for tasks with large datasets and categories \cite{xi2017capsule,phaye2018dense}.
Once all the iterations complete, the features of L capsules should have already been routed to the H capsules and ready to proceed to the following layers.

\textbf{Summary.} Generally, the routing algorithms (e.g., Dynamic Routing, Expectation-Maximization Routing) share the similar execution pattern and exhibit several core features in CapsNet routing procedure: (1) The execution of the RP exhibits strong data dependency and needs to be sequentially processed. (2) The procedure adopts all-to-all computation, which routes all the L capsules to the H capsules and forms aggregation in all possible dimensions. (3) The procedure produces a large amount of intermediate variables. (4) The routing procedure is iteratively processed to generate the dynamic coefficients to pass the feature information. We will discuss these features in relation to performance characterization of CapsNet in Sec.\ref{sec:ct}, and our optimizations on Dynamic Routing in the following Sections can be easily applied to other routing algorithms with simple adjustment.

\begin{table}[t]
	\caption{CapsNet Benchmark Configurations}
	\label{tb:benchmark}
	\vspace{-.5em}
	\resizebox{.48\textwidth}{4em}{
	\begin{tabular}{|c|c|c|c|c|c|}
		\hline
		\multirow{2}{*}{\begin{tabular}[c]{@{}c@{}}Net-\\ work\end{tabular}} & \multirow{2}{*}{Dataset}         & \multicolumn{4}{c|}{Configuration} \\ \cline{3-6} 
		&                                  & BS    & L Caps   & H Caps  & Iter  \\ \hline
		Caps-MN1                                                             & \multirow{3}{*}{MNIST\cite{lecun1998gradient}}   & 100   & 1152     & 10      & 3     \\ \cline{1-1} \cline{3-6} 
		Caps-MN2                                                             &                                  & 200   & 1152     & 10      & 3     \\ \cline{1-1} \cline{3-6} 
		Caps-MN3                                                             &                                  & 300   & 1152     & 10      & 3     \\ \hline
		Caps-CF1                                                             & \multirow{3}{*}{CIFAR10\cite{krizhevsky2009learning}} & 100   & 2304     & 11      & 3     \\ \cline{1-1} \cline{3-6} 
		Caps-CF2                                                             &                                  & 100   & 3456     & 11      & 3     \\ \cline{1-1} \cline{3-6} 
		Caps-CF3                                                             &                                  & 100   & 4608     & 11      & 3     \\ \hline
		Caps-EN1                                                             & EMNIST\_Letter\cite{cohen2017emnist}                   & 100   & 1152     & 26      & 3     \\ \hline
		Caps-EN2                                                             & EMNIST\_Balanced\cite{cohen2017emnist}                 & 100   & 1152     & 47      & 3     \\ \hline
		Caps-EN3                                                             & EMNIST\_By\_Class\cite{cohen2017emnist}                & 100   & 1152     & 62      & 3     \\ \hline
		Caps-SV1                                                             & \multirow{3}{*}{SVHN\cite{netzer2011reading}}            & 100   & 576      & 10      & 3     \\ \cline{1-1} \cline{3-6} 
		Caps-SV2                                                             &                                  & 100   & 576      & 10      & 6     \\ \cline{1-1} \cline{3-6} 
		Caps-SV3                                                             &                                  & 100   & 576      & 10      & 9     \\ \hline
	\end{tabular}}
    \vspace{-1em}
\end{table}

\vspace{-.9em}
\section{Characterization and Analysis}\label{sec:ct}
While the CapsNet starts gaining popularity from both academia and industry, the performance characterization of CapsNet on modern high-performance platforms is largely neglected. Given that GPU has become the major platform for executing CapsNet due to high computation capability and deep optimization on matrix operations (which CapsNet has a large amount of), we adopt some of the state-of-the-art NVIDIA GPU platforms to conduct a comprehensive characterization towards the execution behaviors of various CapsNets listed in Table \ref{tb:benchmark}. We use 4 different datasets and corresponding 12 CapsNets with CapsNet-MNIST like structure (Sec.\ref{sec:caps_struct}), and different configurations on batch size (BS in Table \ref{tb:benchmark}), L capsules, H capsules and routing iteration number.
These CapsNets' inference are processed via PyTorch framework \cite{paszke2017automatic} with the latest deep learning library (i.e., CuDNN \cite{chetlur2014cudnn}), which already enables the state-of-the-art CNN optimizations \cite{cudnnopt}. 

\vspace{-.5em}
\subsection{Overall Evaluation for CapsNet Inference}
\label{sec:DR-bottleneck}

\begin{figure}[t]
	\centering
	\includegraphics[width = .48\textwidth, height = .7in]{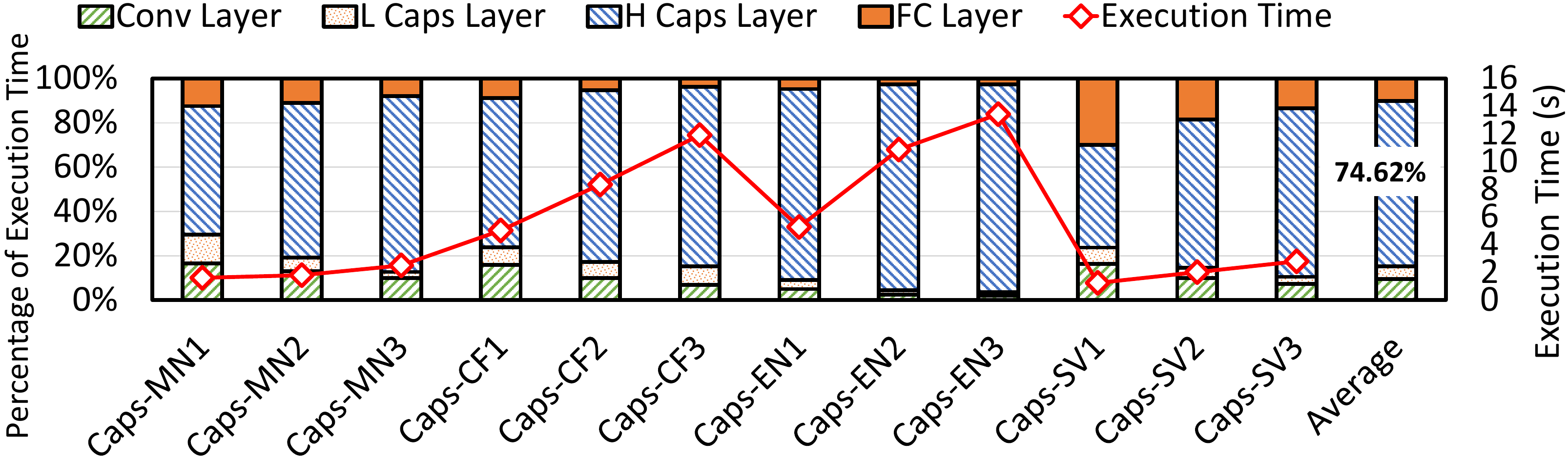}
	\vspace{-1.5em}
	\caption{The overall execution time breakdown of CapsNets on GPU across different layers. Red line represents the actual inference time.}
	\label{fig:PBD}
	\vspace{-.8em}
\end{figure}

Fig.\ref{fig:PBD} demonstrates the performance breakdown of each layer to the overall CapsNet execution. It shows that, across different CapsNet configurations,
the routing procedure (RP) accounts for an average of $74.62\%$ of the entire inference time, becoming the major performance bottleneck. we further conduct detailed analysis on the performance results of CapsNets on GPU and make the following observations:

\vspace{.2em}
\noindent\textit{\textbf{Observation 1: ineffectiveness of batched execution.}} 
One common strategy to accelerate CNNs is to conduct batched execution for improving hardware utilization, especially when input dataset is large. However, it cannot improve the RP's performance during inference. As Fig.\ref{fig:PBD} illustrates, with the increasing of batch size (i.e., Caps-MN1 $\rightarrow$ Caps-MN3),
the overall CapsNet inference time increases; meanwhile, the RP proportion also expands with batch size.

\vspace{.2em}
\noindent\textit{\textbf{\textcolor{black}{Observation 2:  sensitivity to network scaling.}}}
Table \ref{tb:benchmark} shows the network size (formed by a combination of L capsules, H capsules and routing iterations) for each individual case, e.g., Caps-SV1 being the smallest. The red curve in Fig.\ref{fig:PBD} also demonstrates that the overall inference time and RP's percentage generally increases when scaling up the network size (e.g., comparing Caps-MN1, Caps-CF1, Caps-EN1 and Caps-SV1). This implies that RP's execution time is sensitive to network size as well.  

\vspace{.2em}
To summarize, using the highly optimized deep learning library, the RP execution time on GPU can not be effectively reduced through the general CNN optimization techniques such as batch execution. Moreover, it exhibits a certain level of sensitivity to network scaling. Both of these factors make the RP execution a dominating performance bottleneck for CapsNet inference, especially with a growing size and complexity of future CapsNet structures\cite{xi2017capsule,phaye2018dense}.

\vspace{-.5em}
\subsection{Root Causes for Inefficient RP Execution}\label{sec:challenge}

To understand the root causes of RP's inefficiency on GPU, we use NVprofiler \cite{NVProfiler} to collect runtime GPU stats for comprehensive analysis. We observe two root causes for poor RP execution efficiency on GPU-like architectures:

\begin{figure}[t]
	\centering
	\includegraphics[width = .48\textwidth, height = .8in]{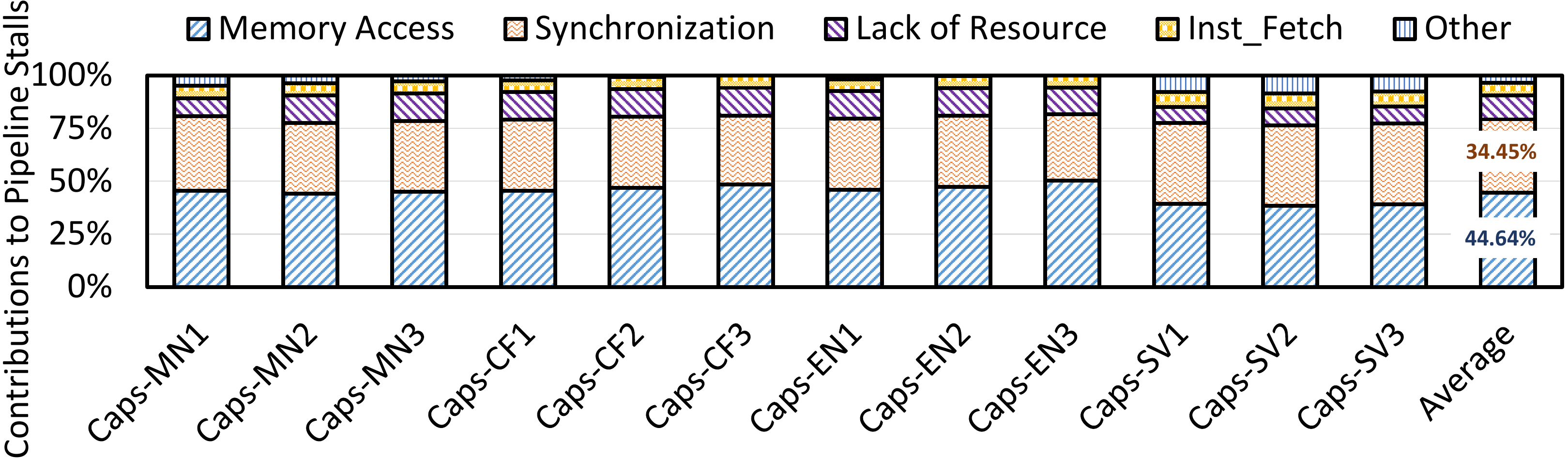}
	\vspace{-1.5em}
	\caption{The breakdown for pipeline stall cycles during RP execution on Tesla P100 Pascal GPU.}
	\label{fig:stall}
	\vspace{-1.5em}
\end{figure}

\begin{figure}[t]
	\centering
	\subfloat[]{\includegraphics[width = .48\textwidth, height = .6in]{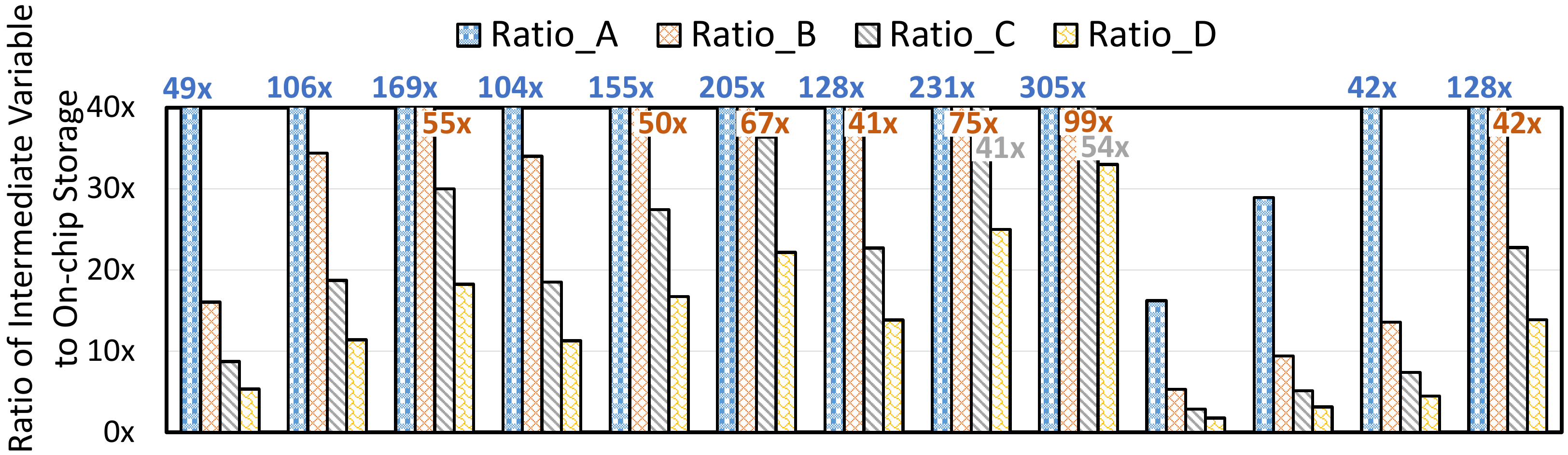}}\\
	\vspace{-1.5em}
	\subfloat[]{\includegraphics[width = .48\textwidth, height = .8in]{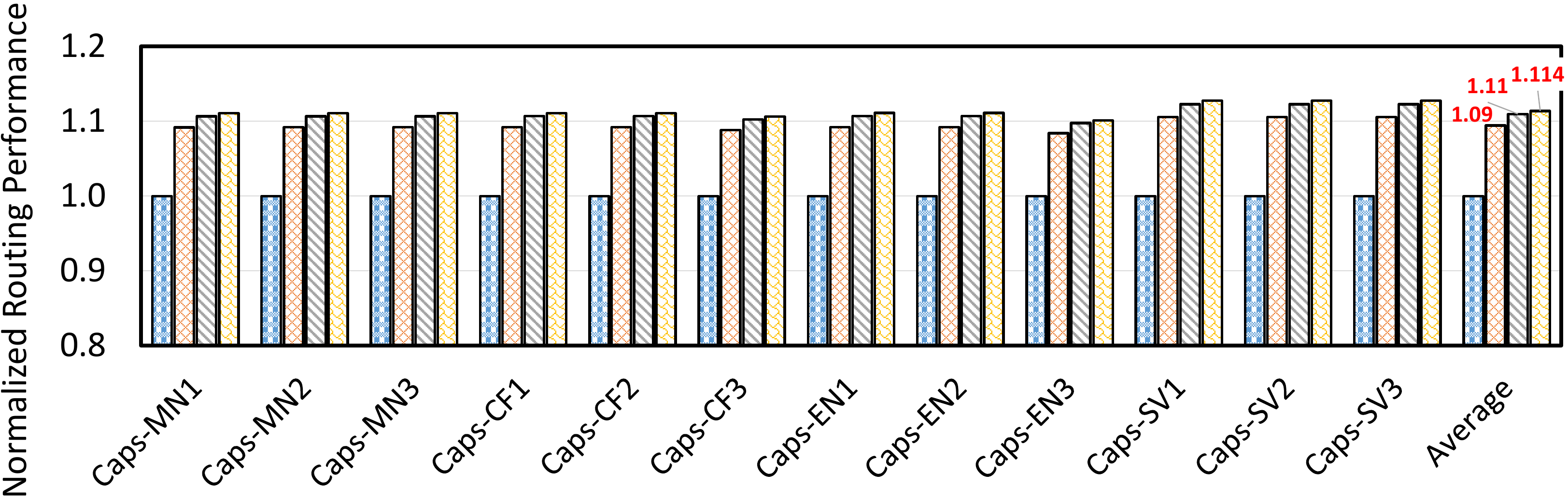}}
	\vspace{-1em}
	\caption{(a) Ratio of intermediate variables' size to on-chip storage of different GPUs; (b) the impact of on-chip storage sizes of state-of-the-art GPUs on RP's execution. A: 1.73MB (K40m), B: 5.31MB (Tesla P100), C: 9.75MB (RTX2080Ti), D: 16MB (Tesla V100).}
	\label{fig:st_ratio}
	\vspace{-1.0em}
\end{figure}

\begin{figure}[t]
	\centering
	\includegraphics[width = .45\textwidth, height = .7in]{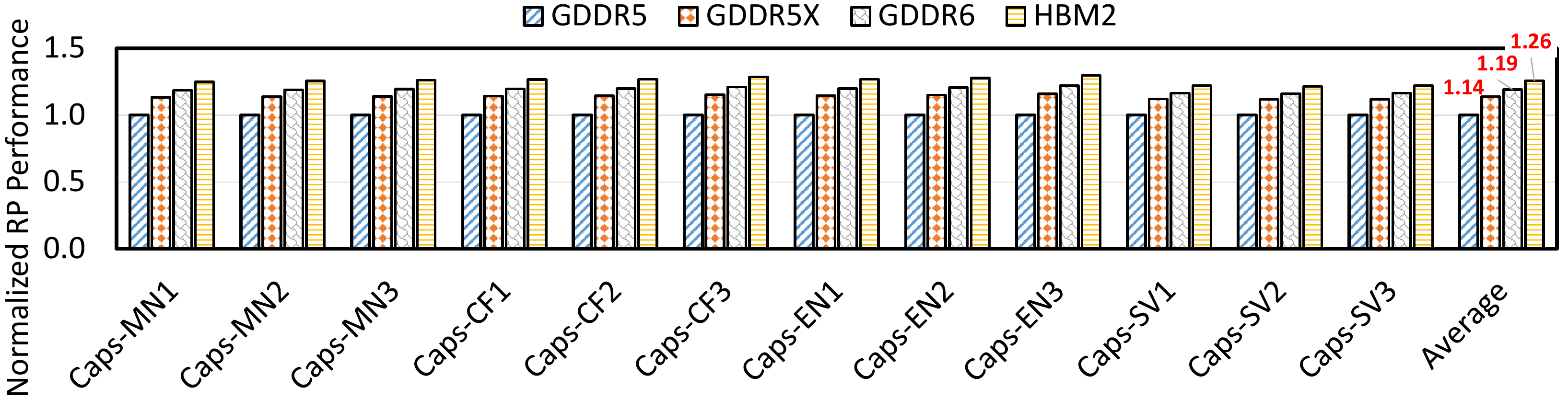}
	\vspace{-.8em}
	\caption{ The impact of memory bandwidth on the overall RP performance. GDDR5:288GB/s (K40m), GDDR5X: 484GB/s (GTX 1080Ti), GDDR6: 616GB/s (RTX 2080Ti), HBM2: 897GB/s (Tesla V100)}
	\label{fig:bw_impact}
	\vspace{-1.3em}
\end{figure}

\begin{figure*}
	\begin{minipage}{.60\textwidth}
		\centering
		\includegraphics[width=\textwidth,height=.95in]{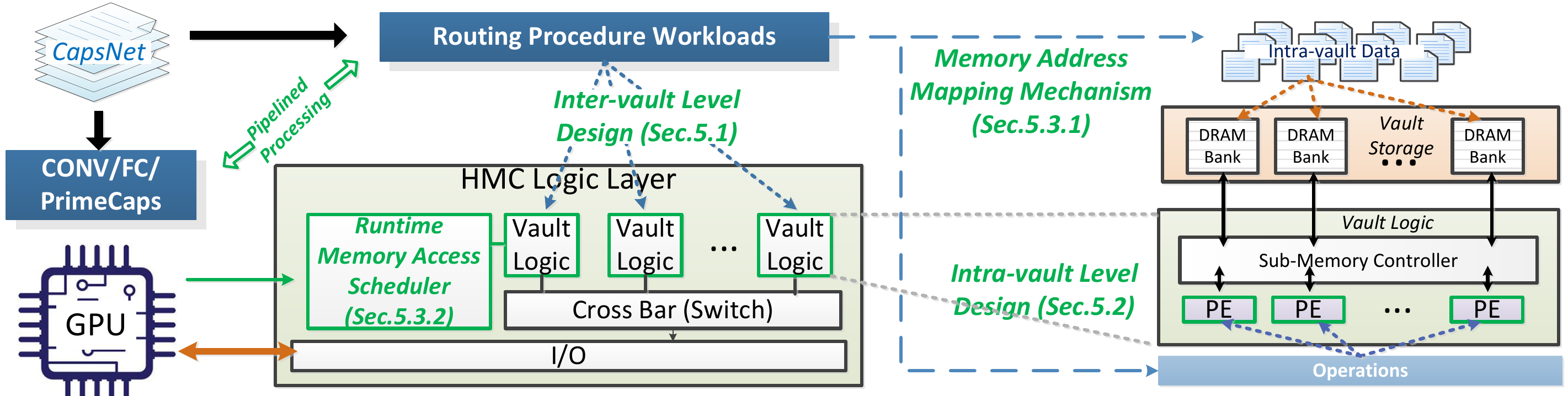}
		\vspace{-1.5em}
		\caption{The overview of PIM-CapsNet Design.}
		\label{fig:pim-caps}
	\end{minipage}
	\quad
	\begin{minipage}{.37\textwidth}
		\centering
		\includegraphics[width = \textwidth,height= .75in]{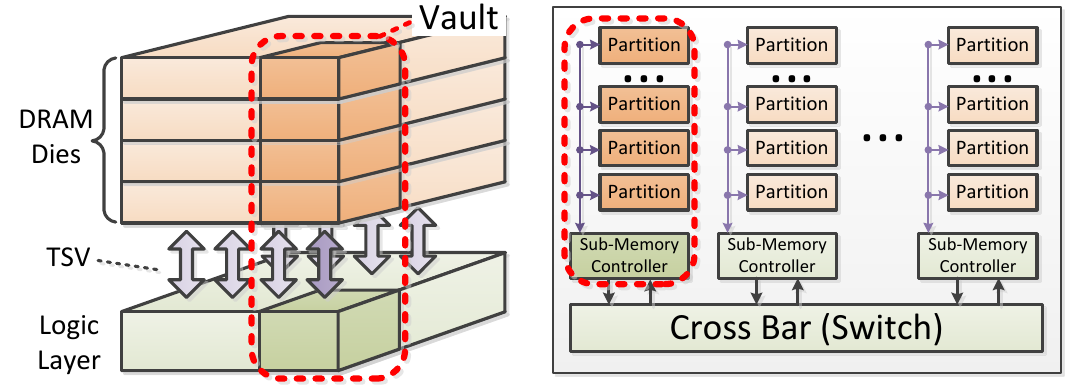}
		\vspace{-1.5em}
		\caption{Left: HMC Organization. Right: HMC Block Diagram. The Red dashed box marks the vault structure.}
		\label{fig:hmc}
	\end{minipage}
	\vspace{-1.5em}
\end{figure*}

\vspace{.5em}
\noindent\textit{\textbf{(1) Significant Off-Chip Memory Accesses:}}
We profile the utilization of several major GPU function units during RP execution on a NVIDIA Tesla P100 GPU. We observe that the arithmetic logic unit (ALU) is lightly utilized (i.e., on average only 38.6\% across the investigated benchmarks) while the load/store unit (LDST) is heavily stressed with an average utilization of 85.9\%. This implies that CapsNets' RP phase fails to effectively leverage the GPU strong computation capability and is severely limited by the intensive off-chip memory access. We further investigate the factors that may contribute to GPU pipeline stalls during RP, including the off-chip memory access, barrier synchronization, lack of resource, etc. Fig.\ref{fig:stall} profiles the individual contribution of each major factor to the overall pipeline stall cycles. As can be seen, the majority of the pipeline stalls are induced by the memory access (i.e., on average $44.64\%$). This further confirms that RP performance is significantly limited by the off-chip memory access. This is caused by a combination of massive intermediate variables from RP execution and limited on-chip storage.
Fig.\ref{fig:st_ratio}(a) illustrates the ratio of RP's intermediate variables' size to on-chip memory sizes of different generations of NVIDIA GPUs. As can be seen, the size of RP's intermediate variables far exceeds the GPU on-chip storage. Given the iterative computation pattern of RP, these variables (e.g., $\hat{u}, s_j, v_j, b_{ij}, c_{ij}$) need to be periodically loaded into GPU cores from the off-chip memory due to the limited on-chip storage. Moreover, the intermediate variables are not sharable among different input batches, which also explains the ineffectiveness of batched execution as we observed in Sec.\ref{sec:DR-bottleneck}. Due to the large data volume and lack of temporal value similarity from these variables, software-level schemes such as register manipulation and shared memory multiplexing are also not very effective.    

\vspace{.2em}
\noindent\textit{\textbf{(2) Intensive Synchronization:}}
Fig.\ref{fig:stall} also indicates that the frequent barrier synchronization is the second major contributor (i.e., on average $34.45\%$) to the pipeline stalls. 
These synchronization overheads are induced by the \textit{\_syncthread()} calls, which coordinate shared memory accesses for all threads in one thread block.
There are two major factors causing the frequent synchronization during the RP execution:
(i) the RP execution contains numerous aggregation operations (e.g., Eq.(\ref{eq:sj})), inducing massive data communication between the threads through shared memory; (ii) the size of the intermediate variables far exceeds the shared memory size, leading to frequent data loading from the global memory. Thus, \textit{\_syncthread()} calls occur frequently to avoid the potential write-after-write and write-after-read hazards.

To address these issues above, we attempt to apply two naive solutions: scaling up the on-chip and off-chip memory capacity. Fig.\ref{fig:st_ratio}(b) shows the impact of on-chip memory sizes of different generations of NVIDIA GPUs on RP execution. We can observe that increasing on-chip memory size can help alleviate the challenges above but not very effective, e.g., only up to an average of 14\% performance improvement for the 16MB V100. This is because the nonsharable intermediate variables' size from RP still far exceeds the current GPUs' on-chip storage, shown in Fig.\ref{fig:st_ratio}(a). Similarly, Fig.\ref{fig:bw_impact} shows that only increasing memory bandwidth from 288GB/s GDDR5 to 897 GB/s HBM slightly improves the overall RP's performance by an average of 26\%. This indicates that higher off-chip memory bandwidth can only solve a small part of the problem but itself does not reduce the high intensity of the off-chip memory accesses. Therefore, to significantly improve CapsNet's inference performance, we need a customized solution to address the root causes for RP's inefficient execution.


\vspace{-1em}
\section{Basic Idea: Processing-in-Memory + Pipelining}

As discussed previously, we aim to address CapsNets' significant off chip-memory access caused by massive unshareable intermediate variables and intensive synchronization due to numerous aggregation operations in RP. Meanwhile, we also want to utilize the excellent core computing capability provided by modern GPUs for deep learning's matrix operations. Thus, we propose a hybrid computing engine named \textit{``PIM-CapsNet"}, shown in Fig.\ref{fig:pim-caps}. It utilizes GPU's native on-chip units as the host for fast processing layers such as Conv and FC, while pipelining with an off-chip in-memory acceleration that effectively tackles RP's inefficient execution. For example, since multiple input sets are generally batched together to be concurrently processed in RP to avoid the local optimal solution of the routing coefficients \cite{mukhometzianov2018capsnet}, host processors can start processing Conv/FC operations from the different batches of the input sets while waiting for RP's results from in-memory processing on the current batch, forming an execution pipeline. Furthermore, the in-memory accelerators of our PIM-CapsNet design can hierarchically improve RP's execution efficiency by minimizing data movement and maximizing parallel processing. Note that our proposed design is a general optimization solution that is applicable to different routing algorithms used in RP.    

To build our in-memory acceleration capability for RP, we resort to one of the emerging 3D stacked technologies, i.e., hybrid memory cube (HMC)\cite{jeddeloh2012hybrid}, which has become a promising PIM platform \cite{yitbarek2016exploring,ahn2015pim,ahn2016scalable,liu2018processing,kim2016neurocube}. The major reason to replace current GPU's off-chip memory (e.g., GDDRX or HBMs) with HMC in our design is their lack of logic layer for in-memory integration. As illustrated in Fig.\ref{fig:hmc}, HMC stacks several DRAM dies on the CMOS logic layer as a cube (specification 2.1 \cite{HMCspec}). The memory cube connects with the host processor (i.e., GPU cores in our case) via the fully-duplex links which can provide up to 320GB/s of external memory bandwidth.
Additionally, the logic layer is split into 32 sub-memory controllers with each communicates its local DRAM banks through Through-Silicon Vias (TSVs) which together provide an internal memory bandwidth of 512GB/s \cite{jeddeloh2012hybrid,ahn2016scalable}.
The sub-memory controller and its local DRAM partitions (each partition contains multiple DRAM banks) form the \textit{vault} architecture, as highlighted in the red dashed box. The logic layer receives system commands and routes memory access to different vaults. A crossbar is integrated in the logic layer to support the communication between SerDes links and vaults. Note that relatively simple computation logic can be integrated onto HMC's logic layer which can directly access data from vaults via memory controllers and benefit from large internal memory bandwidth. This layer is very suitable for integrating in-memory accelerators for RP execution. Next, we will discuss the detailed PIM-CapsNet design.

\vspace{-0.5em}
\section{PIM-CapsNet Architecture Design}

There are two \textbf{key design objectives} in PIM-CapsNet: (i) maintaining workload balance with minimal data communication at \textit{inter-vault level}(Sec.5.1 and Sec.5.3), and (ii) maximizing parallel processing at \textit{intra-vault level} but within architecture design constraints (Sec.5.2 and Sec.5.3). 





\vspace{-.5em}
\subsection{Inter-Vault Level Design} \label{sec:inter}

A typical HMC design adopts crossbar switch to support the inter-vault communication \cite{HMCspec}, and the overall HMC performance power/thermal efficiency can be drastically impacted by the increasing data movements across vaults. Employing a more complicated data flow management (e.g., network-on-chip) on the HMC logical layer would further exacerbate the thermal issue.  
In our proposed design, we leverage the unique features of the RP execution and intelligently distribute workloads across vaults with minimal inter-vault communication.



\subsubsection{RP's Unique Feature: Highly Parallelizable in Multi-Dimensions}

As an interesting feature, the operations of RP equations (Sec.\ref{sec:2.2}) are highly parallelizable. 
For example, in Eq.(\ref{eq:uhat}), the vector-matrix multiplications for all the low- to high-level (L-H) capsule pairs are independent.
We define such independent operations on L capsules as parallelism in the L-dimension, while the independent operations on H capsules are defined as parallelism in the H-dimension.
Additionally, if operations corresponding to different batches are independent, they are defined as parallelism in the B-dimension. Thus, we make the following key observations:

\textbf{Observation I}: \textit{Operations of each equation in the RP can be partitioned into multiple independent sub-operations on at least one of the three dimensions (L, H, or B), suggesting highly parallelizable feature.}


Table \ref{tb:wk} further demonstrates which possible dimensions these five equations of the dynamic routing procedure can be parallelized through. Based on it, we also make the second key observation:

\textbf{Observation II}: \textit{All the RP equations cannot be concurrently executed through the same dimension.}


\begin{table}[t]
	\centering
	\caption{Possible Parallelizable Dimensions}
	\vspace{-.5em}
	\label{tb:wk}
	\resizebox{.45\textwidth}{2.3em}{
	\begin{tabular}{c|ccc}
		\hline
		\textbf{} & \textbf{\begin{tabular}[c]{@{}c@{}}Batch\\ (B-dimension)\end{tabular}} & \textbf{\begin{tabular}[c]{@{}c@{}}Low-level Caps\\ (L-dimension)\end{tabular}} & \textbf{\begin{tabular}[c]{@{}c@{}}High-level Caps\\ (H-dimension)\end{tabular}} \\ \hline
		Eq.\ref{eq:uhat}              & \textbf{x}     & \textbf{x}             & \textbf{x}              \\
		Eq.\ref{eq:sj}              & \textbf{x}     & \textbf{}              & \textbf{x}              \\
		Eq.\ref{eq:vj}              & \textbf{x}     & \textbf{}              & \textbf{x}              \\
		Eq.\ref{eq:bij}              & \textbf{}      & \textbf{x}             & \textbf{x}              \\
		Eq.\ref{eq:cij}              & \textbf{}     & \textbf{x}             & \textbf{}               \\ \hline
	\end{tabular}}
	\vspace{-1.5em}
\end{table}


Observation I indicates that the inter-vault data communication can be reduced by parallelizing the independent sub-operations of each equation on \textit{one chosen dimension} and then distributing them across HMC vaults. By doing so, the major communication across vaults is only required by the aggregation operations at that dimension (aggregations required by the other two dimensions will be performed locally within vaults), which is relatively low.
Observation II, however, emphasizes that the inter-vault communication can not be fully eliminated for RP as none of the three dimensions can support the parallelization for all the RP equations. When distributing the entire RP workloads on a certain dimension, some related data have to be reloaded to another designated vault for aggregation operations.

\vspace{-.5em}
\subsubsection{Parallelization and Workload Distribution}
Since distributing the workloads on different dimensions leads to different communication overheads and execution patterns, it is important to apply an intelligent workload distributor to achieve the optimal design for power and performance. 
To minimize the inter-vault communication, \textit{our distributor only distributes the RP workload on a single chosen dimension.}
Fig.\ref{fig:inter} illustrates an example of the RP execution flow when distributing on the B-dimension.
As it shows, the RP operations of Eq.\ref{eq:uhat},\ref{eq:sj},\ref{eq:vj} (\hquan{1}\hquan{2}\hquan{3}) exhibit parallelism along the B-dimension. Additionally, the multiplication operations of Eq.\ref{eq:bij} (\hquan{4}) (i.e., $v^k_j\times \hat{u}^k_{j|i}$) can also be parallelized along the B-dimension. These parallelizable workloads can be divided into snippets (i.e. green blocks) and distributed across the HMC vaults. Note that typical CapsNet workloads will generate way more snippets than the number of vaults in the HMC (e.g., up to 32 vaults in HMC Gen3).  

Due to the aggregation requirements on all the dimensions during RP execution, there are always some workloads that cannot be processed in parallel on the selected dimension, leading to workload imbalance.
For instance, the remaining operations of the RP procedure in Fig.\ref{fig:inter} (i.e., the purple blocks), including partial Eq.\ref{eq:bij} (\hquan{5}) and Eq.\ref{eq:cij} (\hquan{6}) cannot be divided into snippets due to the lack of parallelism on the B-dimension.
Additionally, these workloads usually require data to be aggregated in one place, making it hard to utilize all the HMC computation resources. Furthermore, the data size requested by the aggregation operations can be large, which may cause massive inter-vault communication and even the crossbar stalls. To reduce the inter-vault communication and increase hardware utilization, we propose to pre-aggregate the partial aggregation operations inside each vault for the corresponding allocated snippets. For example, when the RP workloads are distributed on B-dimension, the pre-aggregation can be performed for Eq.(\ref{eq:bij}) to combine the $b^k_{ij}$ from the snippets assigned to a specific vault before performing global inter-vault aggregation.


\begin{table}[t]
	\centering
	\caption{Parameters for Modeling Inter-Vault Data Movement}
	\label{tb:notation}
	\vspace{-.5em}
	\resizebox{.45\textwidth}{5em}{
	\begin{tabular}{|c|l|}
		\hline
		\textbf{Symbol} & \textbf{Description}                                              \\ \hline \hline
		$I$				& DR iteration number \\ \hline
		$N_B$              & \begin{tabular}[c]{@{}l@{}}Scale for B-dimension\\ i.e. batch size\end{tabular}                        \\ \hline
		$N_L$              & \begin{tabular}[c]{@{}l@{}}Scale for L-dimension\\ i.e. number of low-level capsules\end{tabular} \\ \hline
		$N_H$              & \begin{tabular}[c]{@{}l@{}}Scale for H-dimension\\ i.e. number of high-level capsules\end{tabular} \\ \hline
		$N_{vault}$          & Number of Vault                                                   \\ \hline
		$C_{L}$ 			& \begin{tabular}[c]{@{}l@{}}Dimension of low-level capsule\\ i.e. number of scaler per low-level capsule\end{tabular} \\ \hline
		$C_{H}$ 			& \begin{tabular}[c]{@{}l@{}}Dimension of high-level capsule\\ i.e. number of scaler per high-level capsule\end{tabular} \\ \hline
		$SIZE_{x}$          & Data size of a variable or packet head\&tail                               \\ \hline
	\end{tabular}}
	\vspace{-1.5em}
\end{table}


\vspace{.2em}
\noindent\textit{\textbf{Guiding Distribution via an Execution Score:}}
To achieve the optimal power/thermal and performance results, we propose a metric called execution score $S$ to guide workload distribution, which quantitatively estimates the RP execution efficiency under a given workload distribution. $S$ considers workload assignment to each vault, inter-vault communication overheads, device-dependent factors and inter-vault memory bandwidth. $S$ is modeled as 
$S=1/(\alpha E + \beta M)$.

where $E$ represents the largest workloads distributed to a single vault (since even distribution across vaults is typically not feasible), which can be quantified based on the amount of allocated operations.
$M$ represents the amount of inter-vault data movement. Both $E$ and $M$ are affected by the distribution strategy and the model configuration. Finally, $\alpha$ and $\beta$ represent the device-dependent coefficients, determined by HMC frequency and inter-vault memory bandwidth, respectively.

The calculation of $S$ is independent of the online parameters, thus, the distribution strategy can be determined off-line before the actual inference. Then, the in-vault operations according to this selected distribution dimension will be generated by compiler and the corresponding workloads will be assigned into each vault via a hardware scheduler at runtime. We now demonstrate how to model $S$ via estimating $E$ and $M$ on the three distribution dimensions.

\vspace{.2em}
\noindent\textit{\textbf{Distribution on B-dimension:}}
As Fig.\ref{fig:inter} illustrates, the largest workload assigned to a single vault ($E$) consist of the workload snippets including \hquan{1}\hquan{2}\hquan{3}\hquan{4} and the partial operations \hquan{5}\hquan{6}. With our optimizations, the single vault can get at most $\frac{\lceil\log_{2}(N_{vault})\rceil}{N_{vault}}$ of the unparallelizable operations, where $N_{vault}$ represents number of the HMC vaults.
Using parameters shown in Table \ref{tb:notation}, $E$ can be modeled as follows:
\vspace{-.2em}
\begin{equation}
\begin{aligned}
\resizebox{.9\hsize}{!}{$E_{B}=\lceil\frac{N_{B}}{N_{vault}}\rceil\times N_{L}\times N_{H}\times C_{H}\times (2C_{L}-1) + I\times[\lceil\frac{N_{B}}{N_{vault}}\rceil\times$} \\ 
\resizebox{.8\hsize}{!}{$N_{H}\times C_{H}\times (2N_{L}-1) + 
\lceil\frac{N_{B}}{N_{vault}}\rceil \times N_{H}\times (3C_{H}+19) +$}\\
\resizebox{.8\hsize}{!}{$\lceil\frac{N_{B}}{N_{vault}}\rceil\times N_{L}\times N_{H}\times (2C_{H}-1) + \frac{\lceil\log_{2}(N_{vault})\rceil}{N_{vault}} + 4 \times C_H]$}
\end{aligned}
\vspace{-.1em}
\end{equation}
Since $N_{L}\gg1$, the above equation can be simplified as:
\vspace{-1em}
\begin{equation}
\resizebox{.8\hsize}{!}{$E_{B}=\lceil\frac{N_{B}}{N_{vault}}\rceil\times N_{L}\times N_{H}\times [(4I-1)C_{H}+2C_{L}C_{H}-I]$}
\end{equation}

\vspace{-.2em}
The inter-vault data communication consists of sending pre-aggregated $b_{ij}$ from all the vaults to a single vault and scattering $c_{ij}$ across all the vaults. 
The data transmission is in the form of packets with the head and tail size represented as $SIZE_{pkt}$.
Therefore, the amount of data movements $M$ can be represented as:

\begin{figure}[t]
	\centering
	\includegraphics[width=.48\textwidth,height=.7in]{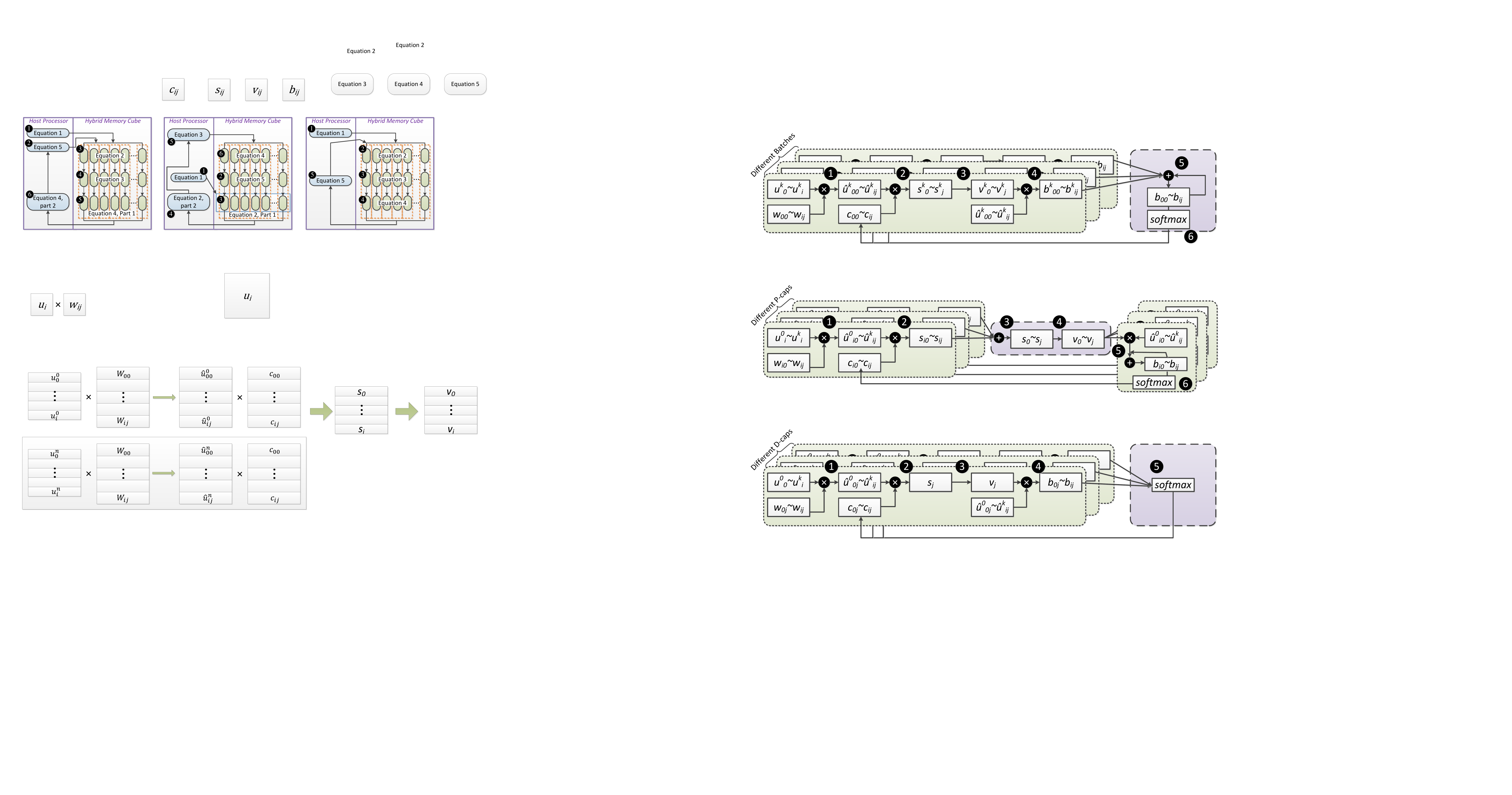}
	\vspace{-1.5em}
	\caption{The execution diagram for the RP procedure with B-dimension distribution. The workloads in green blocks can be split  across vaults, but workloads in purple blocks cannot be distributed via B-dimension.}
	\label{fig:inter}
	\vspace{-1.3em}
\end{figure}

\vspace{-1em}
\begin{equation}
\begin{aligned}
\resizebox{.8\hsize}{!}{$
M_{B} = I\times[(N_{vault}-1) \times N_{L} \times N_{H} \times (SIZE_{b_{ij}}+SIZE_{pkt})$}\\ 
\resizebox{.7\hsize}{!}{$+ (N_{vault}-1) \times N_{L} \times N_{H} \times (SIZE_{c_{ij}}+SIZE_{pkt})]$}
\label{eq:b}
\end{aligned}
\vspace{-.2em}
\end{equation}

\noindent\textit{\textbf{Distribution on L-dimension:}} 
As Table \ref{tb:wk} illustrates, the RP operations of Eq.\ref{eq:uhat},\ref{eq:bij},\ref{eq:cij} can be divided into workload snippets on the L-dimension.
Besides, partial operations from Eq.\ref{eq:sj} (i.e.,  $\hat{u}^k_{j|i}\times c_{ij}$) also exhibit parallelism on the L-dimension. 
Thus, $E$ can be represented as:
\begin{equation}
\resizebox{.8\hsize}{!}{$E_{L}=N_{B}\times \lceil\frac{N_{L}}{N_{vault}}\rceil\times N_{H}\times [2I(2C_{H}-1)+C_{H}(2C_{L}-1)]$}
\end{equation}

The inter-vault communication contains data all-reducing for of $s_j$ and broadcasting $v^k_j$: 

\vspace{-1em}
\begin{equation}
\begin{aligned}
\resizebox{.8\hsize}{!}{$M_{L} = I\times[N_{B} \times (N_{valut}-1) \times N_{H} \times (SIZE_{s^k_j}+SIZE_{pkt})$}\\
\resizebox{.7\hsize}{!}{$+ N_{B} \times (N_{vault}-1) \times N_{H} \times (SIZE_{v^k_j}+SIZE_{pkt})]$}
\label{eq:c}
\end{aligned}
\vspace{-.7em}
\end{equation}

\vspace{.2em}
\noindent\textit{\textbf{Distribution on H-dimension:}} 
As Table \ref{tb:wk} presents, only Eq.\ref{eq:cij} cannot be parallelized on this dimension. 
Hence, $E$ can be represented as
\vspace{-.2em}
\begin{equation}
\resizebox{.7\hsize}{!}{$E_{H}=N_{B}\times N_{L}\times \lceil\frac{N_{H}}{N_{vault}}\rceil\times C_{H}\times [2C_{L}-1+2I]$}
\vspace{-.2em}
\end{equation}

The inter-vault communication contains data all-reducing for of $b_{ij}$ and broadcasting $c_{ij}$: 

\vspace{-1em}
\begin{equation}
\begin{aligned}
\resizebox{.7\hsize}{!}{$M_{H} = I\times[(N_{vault}-1) \times N_{L} \times (SIZE_{b_{ij}}+SIZE_{pkt})$}\\ 
\resizebox{.4\hsize}{!}{$+ N_{L} \times (SIZE_{c{ij}}+SIZE_{pkt})]$}
\label{eq:p}
\end{aligned}
\end{equation}

\vspace{-.3em}

\vspace{-.2em}



\vspace{.3em}

\vspace{-.5em}
\subsection{Intra-Vault Level Design}
In this section, we propose the intra-vault level design that effectively processes the sub-operations of each equation that are allocated to a vault.
We target the design for IEEE-754 single precision (FP32) format, which provides sufficient precision range for CapsNet workloads \cite{sabour2017dynamic}. Our design can also fit other data formats with minor modifications.

\subsubsection{Intra-Vault Level Workload Distribution} \label{sec:intra}

In a basic HMC design, the logical layer of each vault contains a sub-memory controller to handle the memory access. In order to conduct RP specific computation, we introduce 16 processing elements (PEs) into each vault. This design overhead has been tested to satisfy both area and thermal constraints for HMC \cite{kim2016neurocube} (see detailed overhead analysis in Sec.\ref{sec:analysis}). These PEs (integrated onto the logic layer) are connected to the sub-memory controller in the vault, shown in Fig.\ref{fig:intra}(left). Note that the number of parallel sub-operations on certain dimension is generally the orders of magnitude higher than the number of vaults in HMC. In other words, many parallel sub-operations will be allocated to the same vault. Hence they can be easily distributed on the same dimension and concurrently processed via the PEs without introducing additional communication overheads. There may exist some extreme cases that the number of parallel sub-operations allocated to the vault is smaller than the number of PEs, leading to low PE utilization. Since most equations in RP can be parallelized on more than one dimension, the workloads can then be distributed along a different dimension which can produce enough parallelism to fully utilize the PE resources.

\vspace{-.3em}
\subsubsection{Customized PE Design}\label{sec:pe}

\begin{figure}
	\centering
	\includegraphics[width = .4\textwidth,height=0.9in]{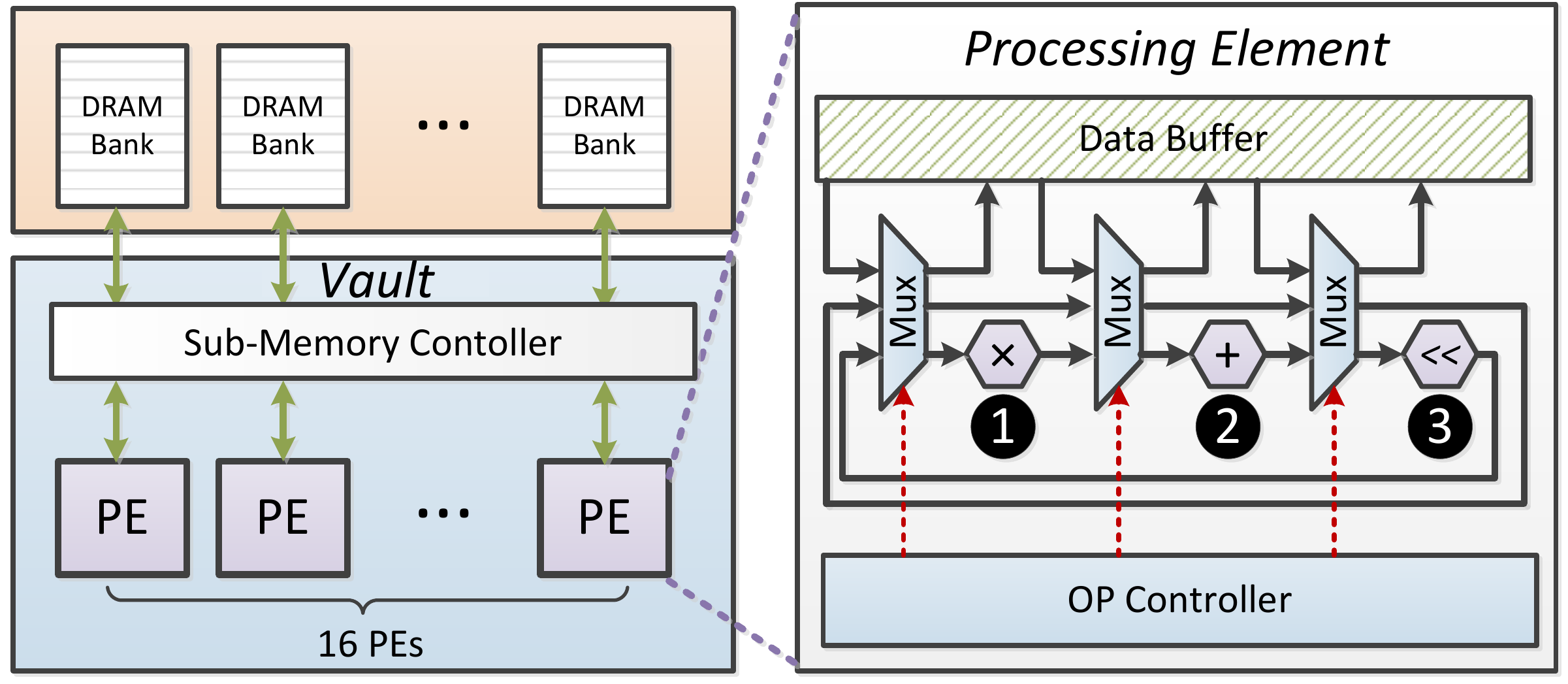}
	\vspace{-.8em}
	\caption{Intra-vault level Architecture Design.}
	\label{fig:intra}
	\vspace{-1.3em}
\end{figure}

\begin{figure*}
	\centering
	\includegraphics[width=.98\textwidth,height= 1.1in]{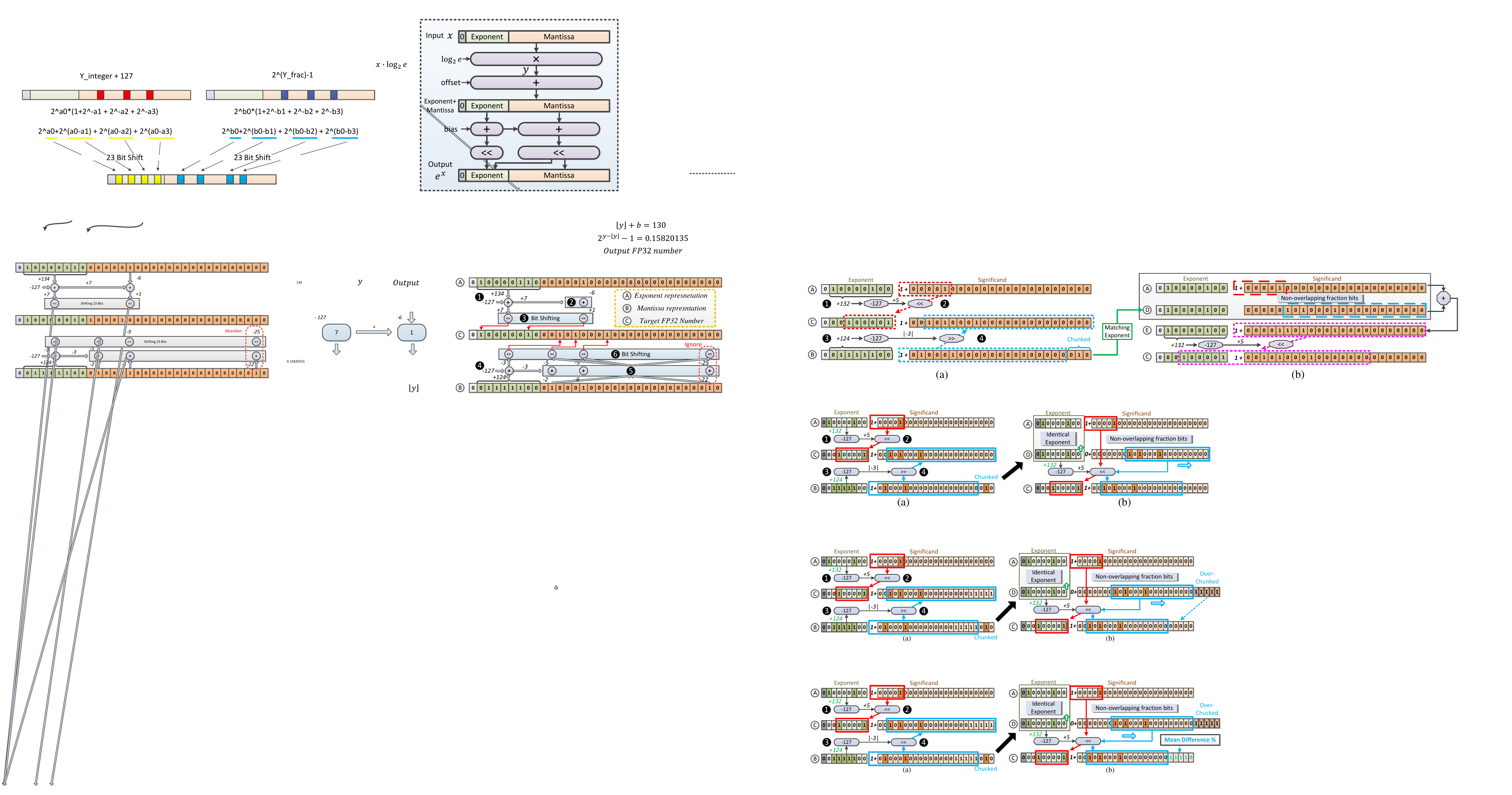}
	\vspace{-1em}
	\caption{(a) An example of transferring the exponent representation \protect\circled{A} (i.e., $\left\lfloor{y}\right\rfloor + b$) and fraction representation \protect\circled{B} (i.e., $2^{y- \left\lfloor{y}\right\rfloor}-1$) to the exponential function's result \protect\circled{C} in FP32 format. (b) Combining the exponent representation \protect\circled{A} and the fraction representation (i.e., \protect\circled{D} that transferred from \protect\circled{B}), and applying a unified bit shifting to obtain the exponential function's result \protect\circled{C}.}
	\label{fig:intra_exp}
	\vspace{-1.5em}
\end{figure*}


There have been some studies \cite{kim2016neurocube,liu2018processing,jeon2018hmc} that integrate adders and multipliers onto the HMC logic layer to perform multiply-accumulation (MAC) which is an essential operation for deep learning (e.g., CNN). 
However, CapsNet's RP execution involves other operations beyond MAC. As discussed in Sec.2.2, among the five RP equations, Eq.\ref{eq:uhat}, Eq.\ref{eq:sj} and Eq.\ref{eq:bij} can be translated into MAC operations.
But the other operations including Eq.\ref{eq:vj} and Eq.\ref{eq:cij} involve more complex functions such as division (Eq.\ref{eq:vj}), inverse square-root (Eq.\ref{eq:vj}) and exponential functions (Eq.\ref{eq:cij}) that require complicated logic design, resulting in large performance and power/thermal overheads \cite{de2009high,partzsch2017fixed}. 

\vspace{.2em}
\noindent{\textit{\textbf{Operation Approximation:}}}
To gain high performance while maintaining low hardware design complexity, we propose approximation to simplify these operations with negligible 
accuracy loss. 
For simplifying division and inverse square-root functions of the FP32, we apply bit shifting \cite{lomont2003fast}, which is widely adopted in graphics processing domain \cite{robertson2012brief,zoglauer2011design,middendorf2013programmable}. For exponential function, we approximate the operation as follows: 

The original exponential function can be transformed 
in the form of power function with the base as 2 \cite{kahan1996ieee}:
\vspace{-.5em}
\begin{equation}\label{eq:exp}
e^x=2^{log_2(e)\times x}=2^y = 2^{\left\lfloor{y}\right\rfloor}(1+2^{y- \left\lfloor{y}\right\rfloor}-1)
\vspace{-.5em}
\end{equation}
where 
$\left\lfloor{y}\right\rfloor$ is the integer part of $y$, and $y- \left\lfloor{y}\right\rfloor$ is the decimal part.


Fig.\ref{fig:intra_exp}(a) illustrates an example of representation transfer.
First, the $ep$ of both A and B will be subtracted from the bias $b$ to get their real exponents (i.e., $ep-b$), as shown in Fig.\ref{fig:intra_exp}(a)\hquan{1}\&\hquan{3}. 
Then, the most significant $ep-b+1$ bits of \circled{A}'s significand (i.e., $1+fraction$) will be chunked and filled into the least significant $ep-b+1$ bits of \circled{C}'s exponent field, with the remaining exponent bits in \circled{C} filled by zeros, as shown in Fig.\ref{fig:intra_exp}(a)\hquan{2}.
We conduct s similar operation to transfer \circled{B} to the \circled{C}'s fraction. Since \circled{B} is a fraction value, its exponent is a non-positive number. \circled{B}'s significand will be logical shift right by $|ep-b|$ bits and then its most significant 23 bits are filled into \circled{C}'s fraction field, as shown in Fig.\ref{fig:intra_exp}(a)\hquan{4}.

The above two transfers can be considered as bit shifting on the significand (i.e., $1+fraction$) in FP32 format with the distance and direction determined by the real exponent (i.e., $ep-b$).
As illustrated in Fig.\ref{fig:intra_exp}, the \circled{B}'s exponent can increase to match the exponent of \circled{A} with its significand bits logically shifting right. By doing this, the faction representation can be described as \circled{D} in Fig.\ref{fig:intra_exp}(b).
Given the matched real exponent value (i.e., $ep-b$), the two representations, i.e., \circled{A} and \circled{D} in Fig.\ref{fig:intra_exp}(b), now share the identical bit shifting operations. Additionally, \circled{A} and \circled{D} correspond to ExpResult's (i.e.,exponential function's result) integer and fraction, respectively. There is no overlapping between their fraction bits. Thus, these two representations can be combined (i.e., \circled{A} OR \circled{D}) followed by a unified bit shifting operation on the significand.
Note that the exponent matching procedure (\circled{B}$\rightarrow$\circled{D}) could over chuck several least significant bits which would originally be mapped into the ExpResult.

Since the exponent matching and combination of two FP32 numbers can be simply considered as a FP32 addition, we can treat the ExpResult computation as an addition of the exponent and fraction representations (i.e., $\left\lfloor{y}\right\rfloor + b + 2^{y- \left\lfloor{y}\right\rfloor}-1$) followed by the bit shifting operations.
Note that the power of 2 function causes high complexity in both execution and logic design, we propose to simplify it as follows:

The above polynomial can be expanded as $(y+2^{y- \left\lfloor{y}\right\rfloor}-(y- \left\lfloor{y}\right\rfloor)+b-1)$; then, the average value $Avg$ of $(2^{y- \left\lfloor{y}\right\rfloor}-(y- \left\lfloor{y}\right\rfloor))$ can be achieved via \textcolor{black}{integrating the polynomial over $(y- \left\lfloor{y}\right\rfloor)\in [0,1)$}, which is fixed and can be obtained offline.
With $y$ represented by $x$, the exponential function can be described as follows:
\vspace{-.5em}
\begin{equation}
\begin{aligned}
\resizebox{.7\hsize}{!}{$ExpResult \simeq BS(log_2(e)\times x+{Avg+b-1})$}
\end{aligned}
\vspace{-.5em}
\end{equation}
where the $BS$ is bit shifting operations with information from $ep-b$; and $log_2{(e)}$ is a constant that is computed offline.

\vspace{.2em}
\noindent\textit{\textbf{Accuracy Recovery:}}
Under the worst case scenari, othere might be several lowest significand bits chucked when mapping from \circled{D} to \circled{C}. It may cause some accuracy loss.
To minimize the bit chucking impact, we analyze 10,000 exponential executions to collect the value differences between the approximated and original results. During the approximation execution, the accuracy loss will be recovered via enlarging the results by the mean percentage of the value difference. Note that our accuracy recovery scheme only introduces one additional multiplication operation during the inference, which guarantees the high performance and the low design complexity compared to other exponential approximation methodology, e.g., applying look-up tables \cite{partzsch2017fixed}.  
Sec.\ref{sec:analysis} shows the detailed accuracy analysis.

\vspace{.2em}
\noindent{\textit{\textbf{Final PE Structure:}}} According to the discussion above, the special functions can be simplified as a combination of addition, multiplication, and bit shifting operations. Thus, our intra-vault PE employs adders, multipliers, and bit-shifters to construct these special functions, as shown in Fig.\ref{fig:intra}. 
Specifically, our PE enables the flow configuration via the multiplexer (MUX) to support different types of operations. For example, PE execution flow \hquan{1}\hquan{2} is for MAC operations; \hquan{3}\hquan{2}\hquan{1}\hquan{2}\hquan{1} is for inverse square-root operations; and \hquan{1}\hquan{2}\hquan{2}\hquan{3} is for exponential function.

\vspace{-.5em}
\subsection{Contention Reduction in CapsNet Design}


In this section, we discuss strategies to combat the memory-level contention in our PIM-CapsNet design. 





\vspace{-.5em}
\subsubsection{Memory Address Mapping Mechanism} \label{sec:cam}

\begin{figure}
	\centering
	\includegraphics[width = .45\textwidth, height=.5in]{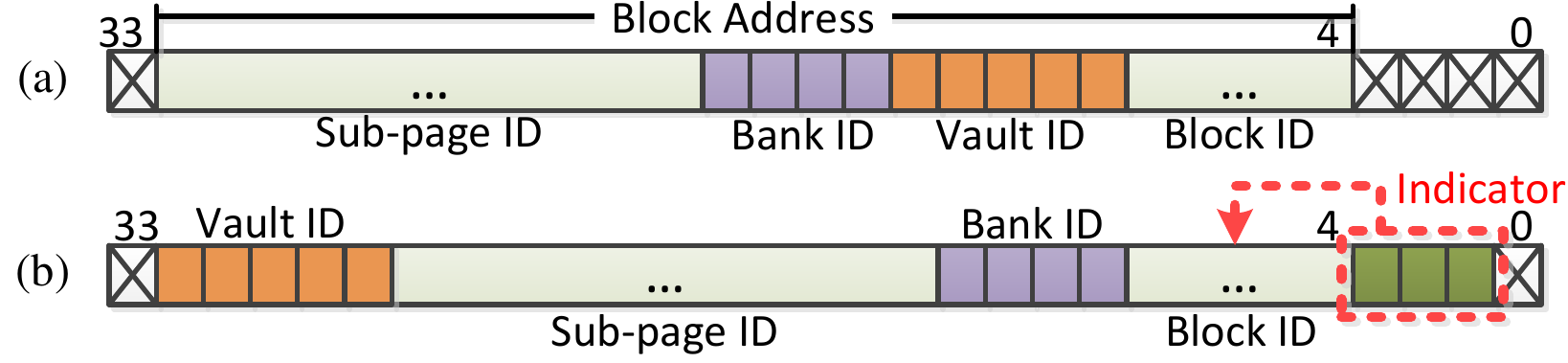}
	\vspace{-1.em}
	\caption{(a)The default address mapping of 8GB in HMC Gen3; (b) Our address mapping.}
	\label{fig:ma}
	\vspace{-1em}
\end{figure}

In the default HMC design, memory access granularity is 16 bytes which is defined as a block, and the sequential interleaving mapping mechanism is applied to benefit from better bandwidth utilization. Moreover, MAX block is introduced to define the maximum number of blocks that one bank can provide at a time. Its size can be set to 32B, 64B, 128B, or 256B \cite{jeddeloh2012hybrid}. In this study, we redefine MAX block as a subpage in order to differentiate it from block, and one subpage is composed of multiple blocks. Fig.\ref{fig:ma}(a) illustrates the default address mapping from HMC Gen3 specifications \cite{jeddeloh2012hybrid}. The lowest 4 bits and the highest 1 bit are ignored, and the remaining bits describe the block address. From its lower to higher bits, a block address is composed of several fields: the block ID in the sub-page (the number of bits is determined by the sub-page size), the 5-bit vault ID for 32 vaults, the 4-bit Bank ID for 16 banks per vault, and the sub-page ID in the bank. As can be seen, sub-pages with consecutive addresses will be first spread sequentially to different vaults and then different DRAM banks. 

Note that our inter-vault level design requires consecutive blocks allocated into one vault to avoid high inter-vault communication. This can be easily addressed by moving up the vault ID field to the highest field of the block address (as shown in Fig.\ref{fig:ma}(b)) so that vault ID remains unchanged 
during the intra-vault memory address mapping. 
However, at the intra-vault level, PEs process their workloads in parallel and concurrently generate data requests, which may result in serious bank conflicts and vault request stalls (VRS).




Interestingly, we observe that most of the concurrent PE requests assigned to the same bank actually visit different blocks. Based on this, we propose a new memory addressing scheme to distribute these blocks to different banks, in order to significantly alleviate bank conflicts and decrease the VRS. 
However, simply distributing blocks to different banks could further increase the VRS as one PE may request multiple consecutive blocks at a time. Because in this case these blocks will reside in multiple banks, it leads to multiple accesses to these banks, resulting in higher bank conflicts. To ensure the consecutive blocks required by one PE are stored in the same bank, our scheme will dynamically determine the sub-page size according to the size of the requested data.
As shown in Fig.\ref{fig:ma}(b), we leverage bit 1 $\sim$ bit 3 in the lowest 4 ignored bits as the indicator to determine the sub-page size for the data requests, where range ``000" $\sim$ ``100" represents the sub-page size from 16B $\sim$ 256B. Given that the data requests from PEs and host GPU need to be allocated into different banks, the indicator bits are dynamically assigned by the page table during the virtual-physical address translation according to the storage requested by each variable involved in each execution thread.


\begin{figure}[t]
	\centering
	\includegraphics[width=.4\textwidth,height=.65in]{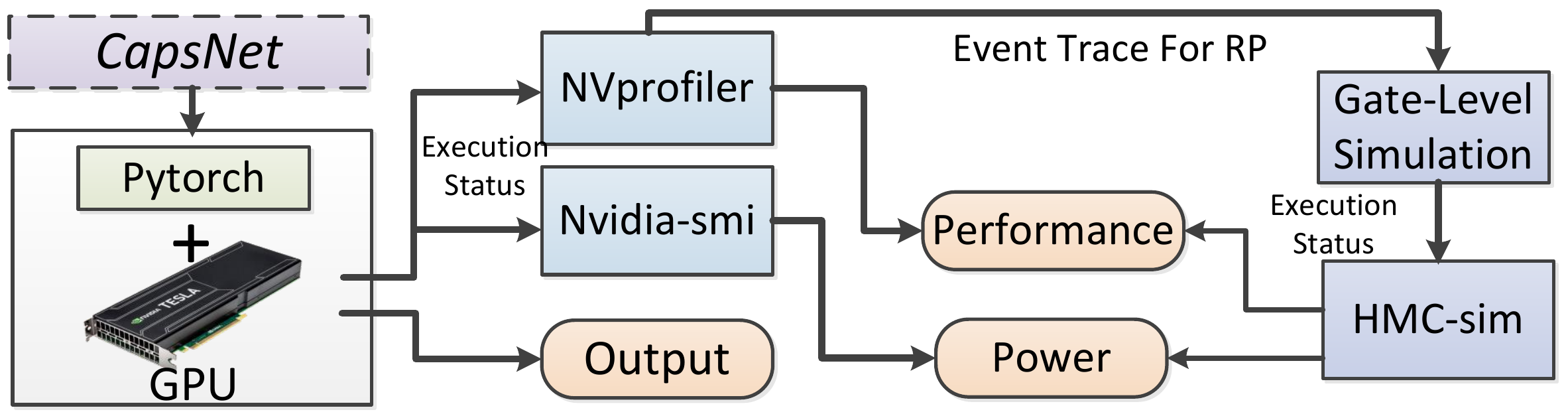}
	\vspace{-1.em}
	\caption{The Evaluation Infrastructure.}
	\label{fig:exp_s}
	\vspace{-1.3em}
\end{figure}

\vspace{-.5em}
\subsubsection{Identifying Memory Access Priority}
During CapsNet execution, resource contention from concurrent data requesting to the same vault from both host GPU and PEs could occur.  
Although the Conv/FC layers exhibit much better data locality than the RP, they may still need to periodically request data from the HMC. 
By default, the priority of a request is determined by the arrival time. But this can cause stalls if both sides are requesting to access the same bank in a vault, which may occur more frequently after applying our new address mapping. 

To address this issue, we propose a runtime memory access scheduler (RMAS) to dynamically identify the priority of memory requests from both the host GPU and vault PEs. We first observe that, with our address mapping mechanism, consecutive data are likely to be distributed across the banks within a vault instead of being scattered over all the vaults.
Thus, it is highly likely that each consecutive data request from the host will only access a single or few vaults at a time instead of all the vaults. This provides opportunities for some vaults to serve the GPU memory requests in rotation without affecting the overall HMC execution.

To quantify the impact of vault access priority, the runtime scheduler (RMAS) first collects the information of the issued operations by HMC and the number of PE requests ($Q$) from the vault request queues in HMC. It also collects the information of the issued operations from the host GPU about which and how many of vaults the operations are requesting from the HMC (defined as $n_{max}$). The collected information above from RMAS is associated with the HMC performance overhead if granting priority to access from either side. Thus, the performance impact of serving the two types of access requests from HMC and GPU can be quantified via the following overhead functions:
\vspace{-.5em}
\begin{equation}
\resizebox{.55\hsize}{!}{$\kappa = \gamma_v \times n_{h} \times \overline{Q} + \gamma_h \times \frac{n_{max}}{n_{h}}$}
\vspace{-.5em}
\end{equation}
where $\kappa$ represents the quantified performance impact;
$\gamma_v$ and $\gamma_h$ represent the impact factors that are determined by the issued operations' type from HMC and host GPU, e.g., memory intensive operations corresponds to a large $\gamma$ as their performance is more sensitive to the memory bandwidth than the computation intensive operations; $\overline{Q}$ is the average number of PEs' requests in the request queues from the targeted vaults;
$n_h$ is the number of vaults that are granted with access priority from the host GPU, which is in the range of $[0,n_{max}]$. If $n_h$ is "0", all the target vaults will first process current HMC PEs requests before processing the GPU requests; while if $n_h$ is $n_{max}$, all the target vaults will grant priority to the GPU requests.
To achieve the minimal impact (i.e. $\min(\kappa)$), $n_h$ should be equal to $\left|\sqrt{\frac{n_{max}\times\gamma_h}{\overline{Q}\times\gamma_v}}\right|$, where $n_h \in [0,n_{max}]$.
The RMAS will then send the control signals to $n_h$ vaults that will give host GPU higher priority to access. Note that a vault with smaller $Q$ has higher priority to be chosen by our RMAS to further reduce the impact on execution.


\vspace{-.5em}
\section{Evaluations}
\vspace{-.5em}
\subsection{Experimental Setup}\label{Sec:ExpSetup}

Fig.\ref{fig:exp_s} shows our evaluation infrastructure to evaluate PIM-CapsNet. We first employ Pytorch \cite{paszke2017automatic}, a popular open-source machine learning framework that supports dynamic computation graphs, to execute the CapsNet on our host GPU. We then design a physical-simulator cooperated platform which takes the execution status of CapsNet provided by Pytorch to obtain the detailed performance and energy results when applying PIM-CapsNet.  
From the physical side, we adopt the Nvidia Tesla P100 \cite{NvidiaP100} as our host processor to evaluate performance and energy consumption of the Conv/PrimeCaps/FC layers of CapsNet. The detailed execution time and power information for these layers are captured by using NVprofiler \cite{NVProfiler} and Nvidia-smi \cite{Nvsmi}. 
From the simulator side, we leverage NVprofile to collect the event trace from host and pass it to a modified HMC-sim 3.0 \cite{leidel2014hmc} to simulate the computing and memory accesses in HMC. Considering that HMC-sim 3.0 cannot provide precise execution cycles and power information for the logic layer design, we conduct a gate-level simulation on Cadence \cite{Cadence} to measure the execution latency and power consumption for our logic design (PE). We then integrate the gate level results and our PIM design in HMC-sim to obtain the performance and energy consumption of RP execution.    
Finally, since the execution of CapsNet is pipelined on the host processor and HMC, we combine the results from both sides via overlapping the execution time and accumulating the energy consumption. The detailed platform configurations are shown in Table \ref{tb:sys}.
In this work, we choose 12 different types of CapsNets as our benchmarks which are introduced in Sec.\ref{sec:ct} and shown in Table \ref{tb:benchmark}.

\begin{table}[t]
	\centering
	\caption{Platform Specifications}
	\vspace{-1em}
	\label{tb:sys}
	\resizebox{.4\textwidth}{4em}{
		\begin{tabular}{|c|c|}
			\hline
			\multicolumn{2}{|c|}{\textbf{Host Processor (GPU)}}                     \\ \hline
			Shading Unit                     & 3584 @ 1190MHz                  \\ \hline
			\multirow{2}{*}{On-chip Storage} & L1Cache/Shared: 24KB x 56      \\ \cline{2-2} 
			& L2 Cache: 4MB               \\ \hline
			Default Memory                   & HBM, 8GB, 320GB/s                    \\ \hline
			\multicolumn{2}{|c|}{\textbf{HMC}}                                \\ \hline
			Capacity                         & 8 GB, 32 Vault, 16 Banks/Vault \\ \hline
			Bandwidth                        & Extl:320 GB/s, Intl: 512GB/s   \\ \hline
			No. of PEs per Vault             & 16                             \\ \hline
			Frequency                        & 312.5MHz                       \\ \hline
		\end{tabular}}
	\vspace{-1.2em}
\end{table}

To evaluate the effectiveness of PIM-CasNet, we compare with the following design scenarios: (1) Baseline: the state-of-the-art GPU accelerated CapsNet execution with the HBM memory (320GB/s). (2) GPU-ICP: the GPU accelerated CapsNet execution with ideal cache replacement policy. 
(3) PIM-CapsNet: our combined inter-vault level and intra-vault level design for RP acceleration with the new memory addressing scheme and RMAS scheme 
(4) PIM-Intra: our PIM-CapsNet without inter-vault level design, and the memory addressing scheme does not optimize the inter-vault data distribution.
(5) PIM-Inter: our PIM-CapsNet without intra-vault level design, and the memory addressing scheme does not support the intra-vault bank conflict optimization.
(6) RMAS-PIM and (7) RMAS-GPU: Our PIM-CapsNet with the naive memory access scheduling, which always grants HMC PEs higher priority than GPU for RMAS-PIM, and always grants GPU higher priority than HMC PEs for RMAS-GPU.
(8) All-in-PIM: the HMC accelerated CapsNet execution, including RP and other layers' execution.

\vspace{-.3em}
\subsection{Effectiveness of PIM-CapsNet}
\vspace{-.5em}
\subsubsection{Performance and Energy for RP execution}\label{sec:vsgpu}

\begin{figure}[t]
	\centering
	\subfloat[]{\includegraphics[width=.45\textwidth,height=.45in]{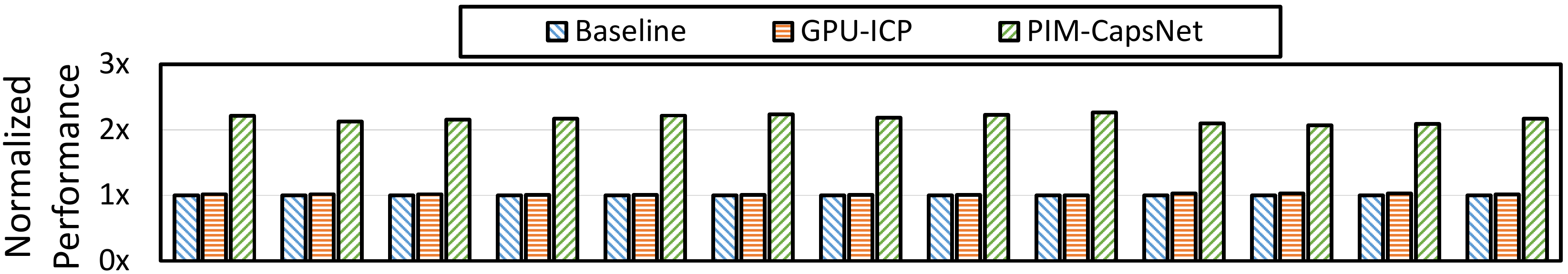}}\\
	\vspace{-1.5em}
	\subfloat[]{\includegraphics[width=.45\textwidth,height=.6in]{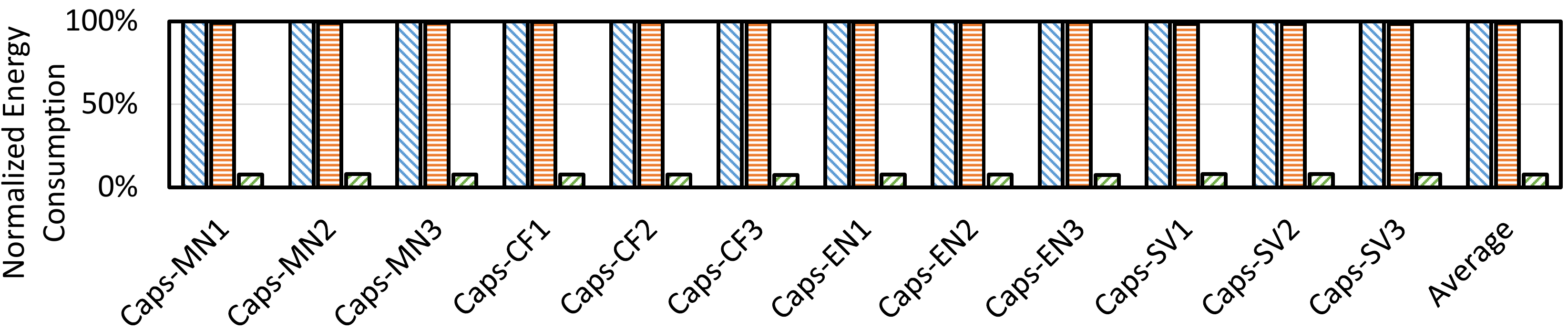}}
	\vspace{-1em}
	\caption{The (a) speedup and (b) normalized energy consumption of PIM-CapsNet on the RP procedure comparing with the GPU-based design.}
	\label{fig:eva_sys}
	\vspace{-1.5em}
\end{figure}

We first evaluate the effectiveness of PIM-CapsNet on RP execution. Fig.\ref{fig:eva_sys} illustrates the normalized performance and energy consumption of our PIM-CapsNet design compared with the GPU-based design (i.e., The Baseline and GPU-ICP) for the RP execution. From the figure, we first observe that
the GPU-ICP only outperforms Baseline by $1.14\%$ on performance and $0.77\%$ on energy during RP execution. This is because RP requires a large number of intermediate variables which exceed the on-chip storage. As a result, the cache policy improvements can barely reduce the off-chip memory accesses.
Second, PIM-CapsNet outperforms Baseline by $117\%$ on performance by addressing the large number of memory access as well as the intensive synchronizations. 
Third, from Fig.\ref{fig:eva_sys}(b), we observe PIM-CapsNet saves $92.18\%$ on energy comsumpting comparing to Baseline. This is because the entire working power of our PIM design is much lower then host and PIM-CapsNet is able to reduce a huge number of data movements between host and HMC.
Moreover, we observe that PIM-CapsNet can achieve better performance and energy saving for RP execution in larger size CapsNet, e.g. $2.27\times$ speedup and $92.52\%$ energy saving of accelerating Caps-EN3 compared with $2.09\times$ speedup and $91.90\%$ energy saving of accelerating Caps-SV1. This implies that PIM-CapsNet exhibits the scalability in optimizing the RP execution with the ever-increasing network size.


\vspace{-.5em}
\subsubsection{Effectiveness of Intra-Vault and Inter-Vault Level Designs}\label{sec:vspim}

\begin{figure}
	\centering
	\subfloat[]{\includegraphics[width=.48\textwidth,height=.6in]{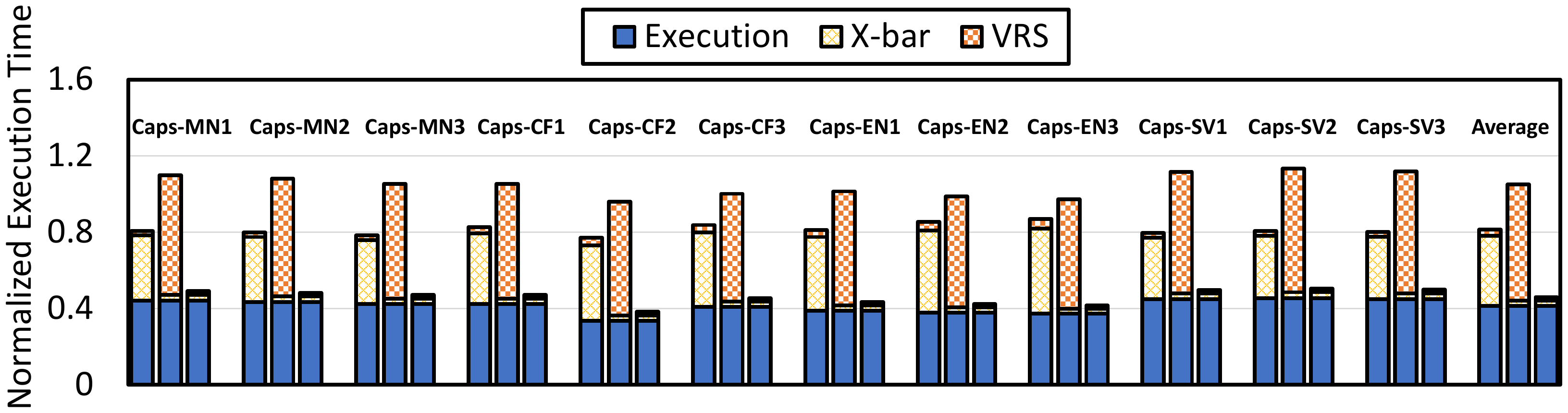}}\vspace{-.5em}\\
	\vspace{-.7em}
	\subfloat[]{\includegraphics[width=.48\textwidth,height=.8in]{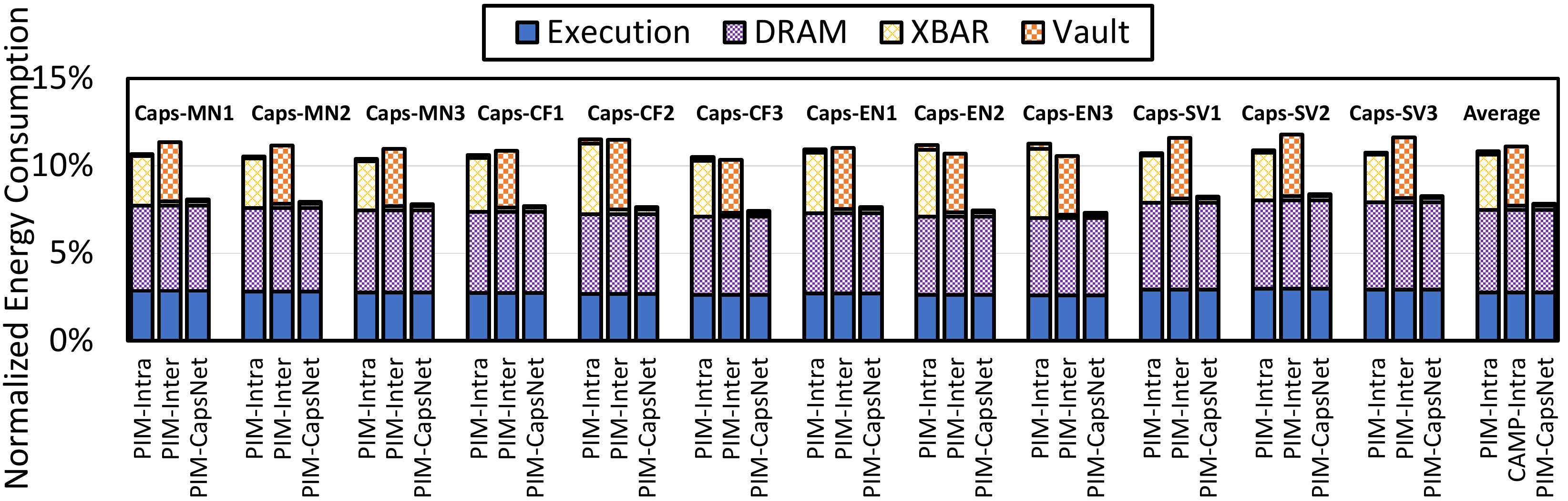}}
	\vspace{-1em}
	\caption{The breakdown of the factor to (a) normalized performance and (b) normalized energy consumption when performing RP execution on different PIM designs.}
	\label{fig:eva_pim}
	\vspace{-1.5em}
\end{figure}

To better understanding the effectiveness of our intra-vault and inter-vault level designs, we compare PIM-CapsNet with other PIM design scenarios for the RP execution only.
Fig.\ref{fig:eva_pim}(a) illustrates the evaluation results of normalized performance with the breakdown factors for different PIM designs. From the figure, we have several observations.
First, even though PIM-Intra achieves $1.22\times$ speedup over Baseline, the inter-vault communication overheads contribute on averages $45.24\%$ to the overall performance.
This is because the PIM-intra design induces massive concurrent data transfer between the centralized computation units in the logic layer and the DRAM banks in vaults, leading to high crossbar utilization and serious stalls.
Second, PIM-Inter decreases the performance by $4.73\%$ compared with the Baseline.
Compared with PIM-Intra, the inter-vault communication overheads have been significantly reduced in PIM-Inter, but the vault request stalls (VRS) grow which contribute on average $57.91\%$ to the execution time due to the serious bank conflicts within the vault. 
Finally, PIM-CapsNet improves the performance about $127.83\%$/$76.62\%$ on average comparing to PIM-Inter/PIM-Intra by reducing both inter-vault communications and VRS.
From the energy perspective, as Fig.\ref{fig:eva_pim}(b) shows, these PIM designs achieve high energy saving compared with Baseline by executing RP on energy-efficient PIM design and our PIM-CapsNet on average outperforms the PIM-Inter/PIM-Intra by $4.81\%$/$4.52\%$ respectively on energy saving. 

\vspace{-.5em}

\subsection{Overall Performance and Energy }

\begin{figure}[t]
	\centering
	\includegraphics[width=.48\textwidth,height=.9in]{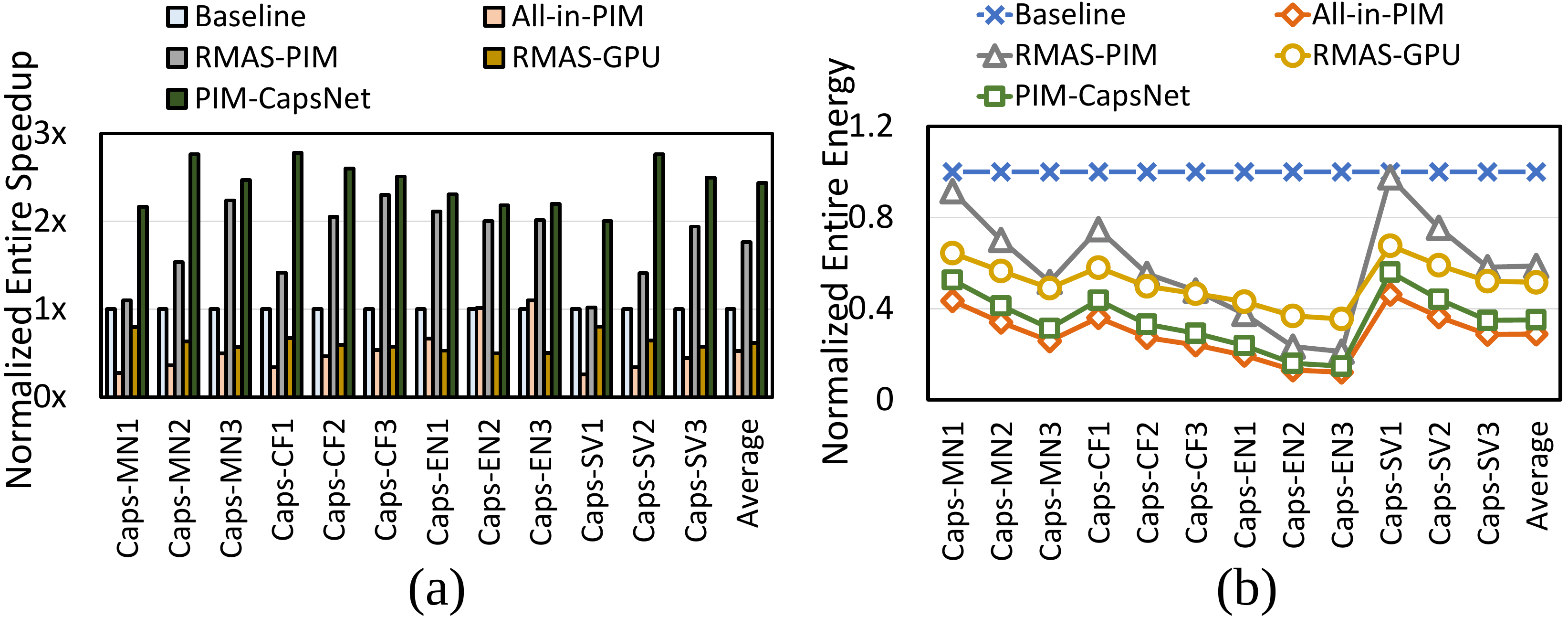}
	\vspace{-1.5em}
	\caption{The (a) speed up and (b) normalized energy consumption when processing entire CapsNet with different designs.}
	\label{fig:eva_overall}
	\vspace{-1em}
\end{figure}

Fig.\ref{fig:eva_overall} shows the normalized speedup and energy of entire CapsNet execution. 
First, to evaluate the effectiveness of the pipelined execution between the host and the HMC, we compare our design with all in GPUs (i.e., Baseline), All-in-PIM.
From the figure, we observe that the All-in-PIM causes the $47.59\%$ performance drop compared to the Baseline. 
This is because we mainly focuses on the RP procedure optimization at minimal cost in HMC, and 
this low-cost design choice can hardly achieve the best performance for Conv/FC layers. But it reduces the energy consumption by $71.09\%$, which exhibits better execution efficiency (i.e. performance /energy consumption) compared with Baseline.
Fig.\ref{fig:eva_overall} Further plots the performance and energy consumption of the pipelined design with native memory access schedulers (i.e., RMAS-PIM and RMAS-GPU). We observe that these naive schedulers can not achieve the best performance and energy saving compared with PIM-CapsNet due to the memory access stalls.
In summary, our design outperforms Baseline on both performance (i.e., on average $2.44\times$) and energy saving (i.e., $64.91\%$), and it even achieves higher performance improvement compared with the acceleration for RP execution only, which benefits from the pipelined execution. 
\vspace{-.5em}
\subsection{Sensitivity to PE Frequency Scaling} \label{sec:freq_scale}

\begin{figure}
	\centering
	\includegraphics[width=.45\textwidth,height=.7in]{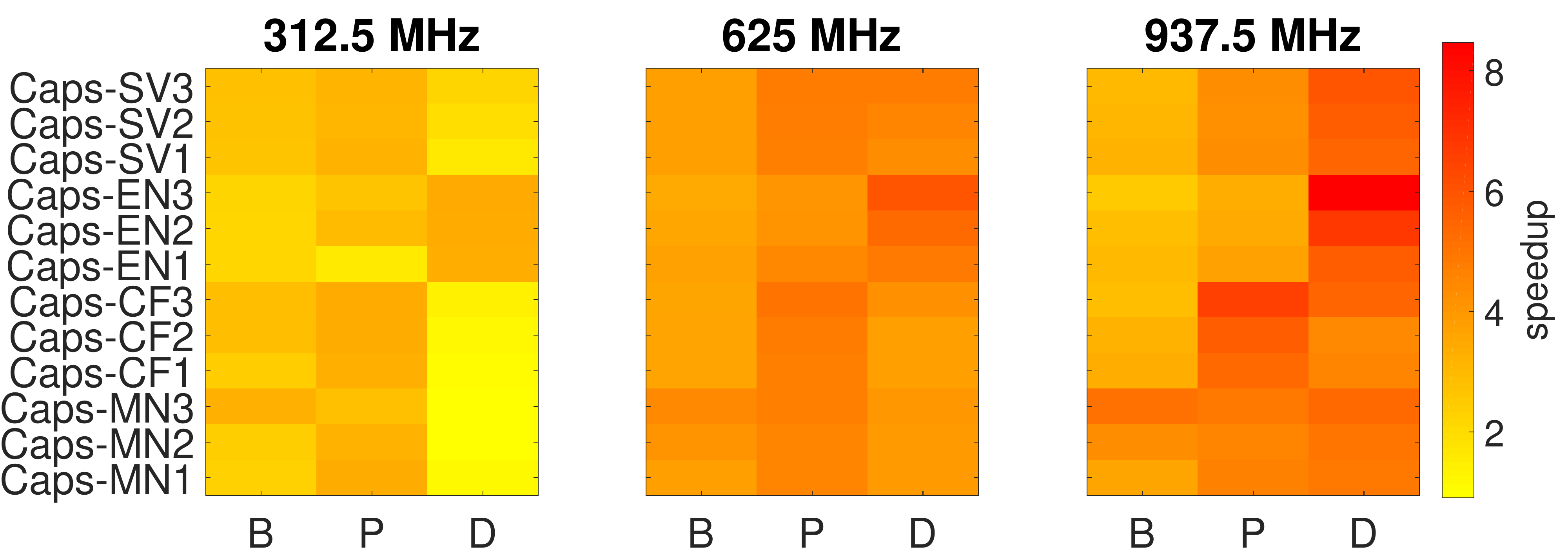}
	\vspace{-1em}
	\caption{The speedup (heat map) achieved by different workloads distribution dimensions (X-axis) under different HMC execution frequency. Red means better improvement.}
	\label{fig:freq_impact}
	\vspace{-1.5em}
\end{figure}

We conduct the sensitivity analysis on inter-vault level workload distributions in our PIM-CapsNet when different frequency is adopted in the PEs, e.g., 312.5MHz, 625MHz and 937.5MHz. Note the frequency scaling will be controlled under a certain range without violating the HMC thermal constraint. 
Fig.\ref{fig:freq_impact} shows the speedup achieved by selecting different distribution dimensions (i.e., B-dimension, L-dimension, and H-dimension) under the above three different frequencies. The darker color indicates the higher speedups (e.g., the red color indicates the substantial speedups while the yellow color means trivial speedups).

It is obvious that PIM-CapsNet can achieve better improvement with higher execution frequency. 
We also notice that the selection of the distribution dimension changes with frequency scaling.
For example, for Caps-SV3, the L-dimension distribution can achieve the best performance under 312.5MHz execution frequency; but the H-dimension distribution achieves the best performance when frequency increases to 937.5MHz.
This indicates that the dimension selection is affected by not only the network configurations, but also the hardware configurations, e.g. processing frequency.

\begin{table}[t]
	\centering
	\caption{Accuracy Validations for PIM-CapsNet}
	\vspace{-1em}
	\label{tb:accuracy}
	\resizebox{.45\textwidth}{3.7em}{
    	\begin{tabular}{|c|c|c|c|c|c|c|}
    \hline
                                                                     & Caps-MN1 & Caps-MN2 & Caps-MN3 & Caps-CF1 & Caps-CF2 & Caps-CF3 \\ \hline
    Origin                                                           & 99.75\%  & 99.75\%  & 99.75\%  & 89.40\%  & 90.03\%  & 90.43\%  \\ \hline
    \begin{tabular}[c]{@{}c@{}}w/o Accuracy \\ Recovery\end{tabular} & 99.73\%  & 99.74\%  & 99.73\%  & 89.15\%  & 89.91\%  & 90.08\%  \\ \hline
    \begin{tabular}[c]{@{}c@{}}w/ Accuracy \\ Recovery\end{tabular}  & 99.75\%  & 99.75\%  & 99.75\%  & 89.37\%  & 90.02\%  & 90.39\%  \\ \hline
                                                                     & Caps-EN1 & Caps-EN2 & Caps-EN3 & Caps-SV1 & Caps-SV2 & Caps-SV3 \\ \hline
    Origin                                                           & 88.74\%  & 85.01\%  & 82.36\%  & 96.70\%  & 95.90\%  & 95.90\%  \\ \hline
    \begin{tabular}[c]{@{}c@{}}w/o Accuracy\\ Recovery\end{tabular}  & 88.19\%  & 84.16\%  & 81.64\%  & 95.08\%  & 95.92\%  & 95.92\%  \\ \hline
    Recovery                                                         & 88.69\%  & 84.96\%  & 82.34\%  & 96.42\%  & 95.90\%  & 95.90\%  \\ \hline
    \end{tabular}}
	\vspace{-1.5em}
\end{table}

\vspace{-.5em}
\subsection{Overhead Analysis}\label{sec:analysis}
\noindent\textit{\textbf{Accuracy Analysis:}}
Table \ref{tb:accuracy} shows the absolute CapsNet accuracy after applying the approximations and accuracy recovery methodologies on our PIM-CapsNet designs as discussed in Sec.\ref{sec:pe}. As it shows, the approximations cause on average $0.35\%$ accuracy loss while the accuracy recovery method can achieve no accuracy loss for most of the cases, and only induces on average $0.04\%$ accuracy difference across our benchmarks.

We also observe the slight accuracy boost for Caps-SV2 and Caps-SV3 when we increase the RP iteration numbers for PIM-CapsNet without accuracy recovery. This is because the noise induced by the approximation enhances the robustness of feature mapping given the enough RP iterations\cite{koistinen1992kernel}.


\vspace{.2em}
\noindent\textit{\textbf{Area Analysis:}}
Our PIM-CapsNet introduces 16 PEs and operation controller into each HMC vault architecture, with one RMAS module located in the logic layer.
Based on our gate-level simulations, our logic design for 32 vaults and the RMAS together incurs $3.11mm^2$ area overheads under the $24nm$ process technology, which only occupy $0.32\%$ HMC logic surface area.

\vspace{.2em}
\noindent\textit{\textbf{Thermal Analysis:}}
Note that our logic design raises the total power of the HMC, which could cause thermal issues for the 3D stacked memory architecture. We observe the average power overhead of our logic design is $2.24W$, which is below the maximum power overhead (i.e., $10W$ \textcolor{black}{thermal design power (TDP)} \cite{zhang2014top}) the HMC can tolerate.

\vspace{-1em}
\section{Related Works}

\noindent\textit{\textbf{PIM-Based NN Acceleration:}}
There have been multiple studies focus on exploring PIM-based neural network accelerators \cite{chi2016prime,shafiee2016isaac,chen2018regan,mao2018lergan,joardar2019regent,falahati2018origami}. For example, \cite{shafiee2016isaac} employs the ReRAM-based PIM to conduct efficient neural network execution. However, the massive intermediate variables involved in the RP could induce both performance and energy overheads in the ReRAM design for frequent value updating.
Besides, \cite{kim2016neurocube,gao2017tetris,liu2018processing} propose the CNN accelerator design using 3D stack PIM technique.
Since the execution pattern of the RP procedure are far different from CNN layers, previous logic layer designs of 3D stacked memory exhibit low efficiency on the RP execution. 
To our best knowledge, this is the first work that leverages HMC to explore a customized architectural design for efficient CapsNet acceleration. 

\vspace{.2em}
\noindent\textit{\textbf{Workload Distributions:}}
Several studies have explored the workload distribution for efficient neural network execution \cite{shen2017maximizing,guo2017bit,wang2019high,lo2017dynamic,dey2018partitioning}. For example, \cite{wang2019high} proposes to distribute the parallel workloads of convolutional layer among the computation units within a single device; and \cite{dey2018partitioning} splits the convolutional execution and distribute the workloads into multiple devices. Their methods have achieved significant CNN accelerations via greatly improving the resource utilization.
Since the execution of the RP procedure is much more complicated than the convolutional layer and involves strong execution dependence, these studies are not applicable to the CapsNet acceleration.

\vspace{.2em}
\noindent\textit{\textbf{Exponential Approximation:}}
There are also some works on approximation for exponential function \cite{de2009high,partzsch2017fixed,malossi2015fast,perini2018fast}. For example, \cite{partzsch2017fixed} leverages the lookup table to conduct the fast exponential execution, which causes substantial performance and area overheads.
\cite{perini2018fast} accelerates exponential function via software optimizations, which are hard to be implemented via the simple logic design.
In this work, we conduct the efficient acceleration for exponential function (Sec.\ref{sec:pe}) with low design complexity and low power overhead.
\vspace{-.5em}
\section{Conclusions}
In recent years, the CapsNet has outperformed the CNN on the image processing tasks and becomes increasing popular in many areas.
In this study, we propose a processing-in-memory based hybrid computing design called PIM-CapsNet, which leverages both GPU core computing capability and the special features of the hybrid memory cube to effectively process the data-intensive RP operations, achieving substantial improvements on both performance and energy saving. To drastically reduce the identified performance bottlenecks of RP, we propose several memory-level optimizations (e,g., multi-dimensional parallelism selection, intelligent workload distribution, complex logic approximation and new address mapping mechanism) based on RP's program and execution features to enable minimal in-memory communication, maximum parallelization, and low design complexity and overhead. Evaluation results demonstrate that our proposed design achieves significant performance speedup and energy savings over the baseline GPU design, for both the RP phase and overall CapsNet inference. It also achieves good performance scalability with increasing network size.


%
%



\bibliographystyle{ieeetr}
\bibliography{ref}

\end{document}